\documentstyle[nato,numreferences]{crckapb} 

\def\beq{\begin{equation}}
\def\eeq{\end{equation}}
\def\beqa{\begin{eqnarray}}
\def\eeqa{\end{eqnarray}}
\def\e      {$^{-1}$}
\def\ee     {$^{-2}$}
\def\eee    {$^{-3}$}
\def\dis{\displaystyle}
\def\frac#1#2	{{#1\over#2}}	
\def\simgt{\lower.5ex\hbox{$\; \buildrel > \over \sim \;$}}
\def\simlt{\lower.5ex\hbox{$\; \buildrel < \over \sim \;$}}

\def\cc     {${\rm cm^{-3}}$}
\def\kms    {${\rm km~s^{-1}}$}
\def\mkms   {{\rm km~s^{-1}}}
\def\mcm    {{\rm cm}}

\def\sun    {${M_\odot}$}
\def\msun   {{M_\odot}}
\def\mug    {\mu{\rm G}}
\def\ug     {$\mu$G}

\def\alf    {Alfv\'en\ }
\def\bom    {{B_{0\mu}}}
\def\dr     {{\rm DR}}
\def\epsin  {{\epsilon_{\rm in}}}
\def\gi     {{\gamma_i}}
\def\gp     {{\gamma_{\rm p}}}
\def\gpi    {{\gamma_{{\rm p}i}}}
\def\gpw     {{\gamma_{\rm pw}}}
\def\gw     {{\gamma_{\rm w}}}
\def\kpi    {{K_{{\rm p}i}}}
\def\w      {{\rm w}}
\def\kcut   {{k_{\rm cut}}}
\def\kp     {{K_{\rm p}}}
\def\msfc   {M({\rm SFC})}
\def\mucr   {{\mu_{\rm cr}}}
\def\mw     {{M_{\rm w}}}
\def\mbe    {M_{\rm BE}}
\def\mcr    {M_{\rm cr}}
\def\mj     {M_{\rm J}}
\def\mphi   {M_\Phi}
\def\nuni   {{\nu_{ni}}}
\def\nh     {n_{\rm H}}
\def\Nh     {N_{\rm H}}
\def\Nbh    {{\bar N_{\rm H}}}
\def\Nbhtt  {{\bar N_{\rm H22}}}
\def\htwo   {{\rm H}_2}
\def\nhtwo  {n_{\rm H_2}}

\def\phad   {{\phi_{\rm AD}}}
\def\pth    {P_{\rm th}}
\def\rhoc   {\rho_{c}}
\def\rhos   {\rho_{s}}
\def\snt    {\sigma_{\rm nt}}

\def\tad    {{t_{\rm AD}}}
\def\tff    {{t_{\rm ff}}} 
\def\tgs    {{t_{g*}}}
\def\tgsdr  {t_{g*{\rm DR}}}
\def\thirteenco   {$^{13}{\rm CO}$}
\def\vcl    {{V_{\rm cl}}}
\def\vad    {{v_{\rm AD}}}
\def\vrms   {{v_{\rm rms}}}
\def\calg   {{\cal G}}
\def\call   {{\cal L}}
\def\calm   {{\cal M}}
\def\caln   {{\cal N}}
\def\calt   {{\cal T}}
\def\calw   {{\cal W}}
\def\vecB   {{\bf B}}
\def\vecg   {{\bf g}}
\def\vecJ   {{\bf J}}
\def\vecr   {{\bf r}}
\def\vecv   {{\bf v}}
\def\vecnabla	{{\bf \nabla}}
\def\vectimes	{{\bf \times}}

\begin{opening}
\title{THE DYNAMICAL STRUCTURE AND EVOLUTION OF GIANT MOLECULAR CLOUDS}
\author{CHRISTOPHER F. MCKEE}
\institute{Institute for Advanced Study\\
Princeton NJ 08540\\
and\\
Departments of Physics and Astronomy\\
University of California, Berkeley CA 94720\footnote{Permanent address}}
\end{opening}
\runningtitle{STRUCTURE AND EVOLUTION OF GMCS}

\begin{document}

\section{Introduction: The Observed Characteristics of GMCs}
\label{sec:int}

	The interstellar medium (ISM) of galaxies contains
gas that spans a wide range of physical conditions,
from hot X-ray emitting plasma to cold molecular gas.
The molecular gas is of particular importance because it
is believed to be the site of all the star formation
that occurs in galaxies.
In the Milky Way, molecular gas constitutes about half the total
mass of gas within the solar circle.  Much of this gas is
concentrated in large aggregations called giant molecular
clouds (GMCs), which have masses $M\simgt 10^4$ \sun.
Smaller molecular clouds are also observed, such
as the high latitude clouds discovered by Blitz et al.
\cite{bli84} and the small molecular clouds in the Galactic
plane cataloged by Clemens \& Barvainis \cite{cle88}.
GMCs have internal structure, and
I shall follow the terminology of Williams et al.
\cite{wil99} in describing this: {\it Clumps} are coherent regions
in longitude--latitude--velocity space that
are generally identified from spectral line maps
of molecular emission.  {\it Star--forming clumps} are
the massive clumps out of which stellar clusters form.
Finally, {\it cores} are the regions out of which single
stars (or multiple stellar systems like binaries)
are formed.  These characteristics, together
with the important 
observational properties of GMCs, are
reviewed elsewhere in this volume by Blitz.  In this
review I shall first briefly summarize some of the key
properties of GMCs and then attempt to account for the
dynamical properties theoretically.

\subsection{CHEMICAL AND THERMAL PROPERTIES}
\label{sec:che}

	Molecular clouds are composed primarily of $\htwo$,
but this is relatively difficult to observe.  The next most
abundant molecule is generally CO, which can be readily 
observed in both emission and absorption.  It is surprising
that such molecules can exist in the harsh environment
of interstellar space, particularly because of the destructive
effects of ultraviolet radiation.  Atomic hydrogen (H$^0$)
shields most interstellar gas from EUV photons
(those with energies 100~eV$\simgt h\nu\geq 13.6$~eV),
but FUV photons (those with 13.6~eV$> h\nu\simgt 5$~eV) are
far more penetrating.  These photons ionize atoms
such as C, Mg, S, and Fe,
and photodissociate molecules.  The photodissociation
of $\htwo$ and CO
occurs in a two step process: first, the molecule undergoes
an electronic excitation by absorption of a resonance line
photon; then some fraction of 
the molecules radiate into a state in the vibrational
continuum and fly apart \cite{dal87}.  

	A significant column density of molecules can build up
only when the absorption lines become optically thick, or,
if this is inadequate, when the FUV radiation is sufficiently
attenuated by dust. Under typical conditions in the local
ISM, observations show that the extinction to the 
cloud surface must exceed about 0.1 mag
in order for a significant column density
of $\htwo$ to be observed \cite{boh78} (the extinction
through the entire cloud is then twice this, or
about 0.2 mag).  Since CO is less abundant than $\htwo$,
a significantly larger column density is required in
order for the carbon to become incorporated into molecules.
The calculations of van Dishoeck \& Black \cite{van88}
show that a GMC in the local ISM has a layer of C$^+$ and
C$^0$ corresponding to a column density $\Nh=1.4\times
10^{21}$ cm\ee. (Note that we shall measure all densities
in terms of the total hydrogen density, $\nh=2\nhtwo$ for molecular gas,
and similarly for column densities.) For the dust to gas
ratio observed in the local ISM, the relation between 
extinction and hydrogen column is \cite{spi78}
\beq
A_V=\frac{\Nh}{2.0\times 10^{21}~{\rm cm}^{-2}} ,
\label{eq:av}
\eeq 
so this column is equivalent to an extinction of 0.7 mag.

	An interstellar cloud with a mean extinction significantly
greater than $2\times 0.7$~mag is thus expected to have
a thin outer layer of H$^0$ ($\Delta A_V\simeq 0.1$~mag),
a thicker layer of $\htwo$, C$^+$, and C$^0$ 
($\Delta A_V\simeq 0.6$~mag), and an interior that is
nearly fully molecular.  The temperature of the outer atomic
layer is of order 50-100 K.  In the deep interior, where
the gas is fully molecular, the temperature of about 10 K
is set by the balance between
heating due to cosmic ray ionization and cooling due
to CO emission.
Because the chemical and thermal structure
of the edge of the cloud is dominated by photodissociation,
it is termed a photodissociation region \cite{hol99}.

	The ionization in 
most of the volume of a molecular cloud is due to FUV photons;
under typical conditions cosmic rays dominate the ionization 
only for regions in a cloud
such that the extinction to the surface
exceeds about 4 mag \cite{mck89}.  The ionization 
in the cosmic ray ionized region of
a molecular cloud can be expressed as
\beq
x_e\equiv \frac{n_e}{\nh} =\frac{C_i}{\nh^{1/2}} .
\label{eq:xe}
\eeq
Williams et al. \cite{wil98} found $C_i=(0.5-1.0)\times 10^{-5}
$~cm$^{-3/2}$
from chemical modeling of observations of a number of
low mass cores, where the factor
of two uncertainty arises from uncertainties in
the chemical reaction rates.  In arriving at this result,
they estimated that the typical density
in the cores they observed is $\nh\simeq 2-6\times 10^4$ cm\eee\
and that the cosmic ray ionization rate is $\zeta_{\rm H}=
2.5\times 10^{-17}$ s\e.  
This value for the ionization is in good agreement with
theoretical expectations \cite{mck93}.

\subsection{DYNAMICAL PROPERTIES}
\label{sec:dyn}

	Some of the most salient characteristics of GMCs 
were summarized in 1981 by Larson \cite{lar81}, 
and these results are sometimes
referred to as ``Larson's laws".  The first relation
is the {\it line width--size relation}: molecular clouds
are supersonically turbulent with line widths
$\Delta v$ that increase as a power of the size,
$\Delta v\propto R^p$.  Larson himself estimated that
$p\simeq 0.38$, close to the value 1/3 appropriate for
turbulence in incompressible fluids.  Subsequent work
has distinguished between the relation valid for
a collection of GMCs and that valid within individual
GMCs or parts of GMCs.  For different GMCs inside the solar circle,
Solomon et al \cite{sol87} found
\beq
\sigma=(0.72\pm 0.07)R_{\rm pc}^{0.5\pm 0.05}~~~~\mkms,
\eeq
where $\sigma$ is the one--dimensional velocity
dispersion; it is related to the full-width half-maximum
of the line by $\sigma=\Delta v/2.355$.
By comparison, the thermal velocity dispersion
of molecular gas is $\sigma_{\rm th}=0.188(T/10~{\rm K})^{1/2}$
km s\e.
Within low--mass cores,
Caselli \& Myers \cite{cas95} found that the
non--thermal velocity dispersion (i.e, what
remains after eliminating the thermal contribution) is
\beq
\snt\simeq 0.55 R_{\rm pc}^{0.51}~~~~\mkms,
\label{eq:snt}
\eeq
which is quite similar to that found for GMCs by
Solomon et al.  For ``high--mass cores", however,
Caselli \& Myers found
\beq
\snt\simeq 0.77 R_{\rm pc}^{0.21}~~~~\mkms,
\label{eq:snt2}
\eeq
a substantially flatter relation. Although these cores
are forming more massive
stars than the low mass cores, they are
not forming OB stars.  Plume et al \cite{plu97}
have surveyed clumps
in which OB star formation is believed
to be occurring.  They found that such clumps do not
obey the line width--size relation, and furthermore,
that $\snt$ is 
greater than indicated by equation (\ref{eq:snt2}).

	Larson's second result was that GMCs are
gravitationally bound.  We shall discuss this 
in \S \ref{sec:are} below.  He concluded that
clumps within GMCs are also gravitationally bound,
but for \thirteenco\ clumps
this appears to be true only for the most massive
ones \cite{ber92}.

	His third conclusion was that all GMCs have
about the same column density.  As he pointed out,
only two of these three conclusions are independent;
any one of them can be derived from the other two.
For example, if the clouds are gravitationally
bound, then $\sigma^2\propto GM/R\propto \bar\Nh R$,
where $\bar\Nh=M/(\pi\mu_{\rm H} R^2)$ is the mean column density
of the cloud, and $\mu_{\rm H}=2.34\times 10^{-24}$~g is the mean
mass per hydrogen. The line width--size relation
for GMCs, $\sigma^2\propto R$ \cite{sol87} then
implies $\bar\Nh=$~const.  For GMCs inside the solar
circle, Solomon et al \cite{sol87} found
\beq
\Nbhtt=(1.5\pm0.3)R_{\rm pc}^{0.0\pm 0.1}~~~~
	{\rm cm}^{-2},
\eeq
where $\Nbhtt\equiv \Nbh/(10^{22}$ Hydrogen nuclei cm\ee);
this corresponds to an extinction $\bar A_V= 7.5$ mag with the local
gas to dust ratio.  This result does not apply to unbound clumps
in GMCs, which can have lower column
densities \cite{ber92}, nor to the
OB star--forming clumps studied by Plume et al \cite{plu97},
which have $\Nbhtt\sim 60$.

	There are several other characteristics of GMCs that
we must take note of.  First, GMCs appear to have magnetic fields
that are dynamically significant
\cite{hei93}; this will be discussed in \S \ref{sec:mag}
below.  Second, GMCs are highly clumped, in that
the typical density $\nh$---i.e., the local
density around a typical
molecule---is significantly greater than the volume--averaged 
density $\bar\nh$ in the cloud.
Liszt \cite{lis95} has summarized studies of
the typical density of clouds in the Galactic plane as inferred
from their excitation, and cites values of $\nh$ ranging from
$10^3$ to $1.2\times 10^4$~\cc; in other words,
$\nh\simeq 3000$ cm\eee\ $\pm 0.5$ dex.
However, the mean
density of gas in the GMCs, $\bar\nh$, is considerably less,
at least for the massive ones: since $M\propto
\bar\nh R^3$ and $\bar \Nh\propto \bar\nh R$, we have
\beq
\bar\nh=\frac{84}{M_6^{1/2}} \left(\frac{\Nbhtt}{1.5}
	 \right)^{3/2} ~~~~\mcm^{-3},
\label{eq:barnh}
\eeq
where $M_6\equiv M/(10^6~\msun)$.
The filling factor of the gas in GMCs is then typically
\beq
f\equiv\frac{\bar\nh}{\nh} = \frac{0.084}{n_{\rm H3}M_6^{1/2}}
	\left(\frac{\Nbhtt}{1.5}
	\right)^{3/2},
\label{eq:f}
\eeq
where $n_{\rm H3}=\nh/(10^3$~cm\eee).  Clouds of mass
$M\simlt 10^4$ \sun\ must have $\nh>10^3$~cm\eee\ if they
are to have the typical column density found by
Solomon et al \cite{sol87}.  The nature of the low
density interclump medium is uncertain; for example,
it is not even known whether it is atomic or molecular
\cite{wil99}.
A possible explanation for the small
filling factor of the gas in GMCs will be given 
below, where we discuss models of turbulence in these clouds.

	Finally, GMCs have a power-law mass distribution with
a relatively sharp cutoff.
Let $d\caln_c(M)$ be the number of GMCs with mass
between $M$ and $M+dM$.  Then the observations of
GMCs inside the solar circle 
(but excluding the Galactic Center) are consistent with
the mass distribution \cite{wil97}
\beqa
\frac{d\caln_c}{d\ln M}
	& = &63\left(\frac{\dis 6\times 10^6~\msun}{\dis M} 
		\right)^{0.6} ~~~~ (M\leq 6\times 10^6\;\msun), \\
	& = &0 ~~~~~~~~~~~~~~~~~~~~~~~~~~~~~~
	(M>6\times 10^6\; \msun).
\label{eq:calnc}
\eeqa
The mass distribution is often represented in terms of $d\caln_c/dM$
instead of $d\caln_c/d\ln M$, which leads to an exponent
of 1.6 instead of 0.6.  The form adopted here has the 
advantage that $d\caln_c/d\ln M$ represents an actual
number of clouds.
The fact that the coefficient 63 is so much larger than
unity means that it must have a physical significance
\cite{mck97}:
If there were no cutoff to the distribution, one would
expect $63/0.6\simeq 100$ GMCs more massive than $6\times 10^6$ \sun\
in the Galaxy; in fact, there are none.  

\subsection{STAR FORMATION IN GMCS}
\label{sec:star}

	GMCs are the sites of most of the star formation
in the Galaxy.  A crucial fact about star formation is 
that it is usually very inefficient: In the absence of support,
GMCs would collapse to very high densities and presumably
form stars in a free fall time
\beq
t_{\rm ff}=\left(\frac{3\pi}{32G\bar\rho} \right)^{1/2}
       =\frac{1.37\times 10^6}{\bar n_{\rm H3}}^{1/2} ~~~~{\rm yr}.
\eeq
(As we shall see in \S \ref{sec:mod}, simulations indicate
that it is very difficult to maintain
the turbulence that supports GMCs.)  
Now, the total mass of GMCs inside the solar
circle is $10^9\; \msun$ \cite{wil97}, and
the mean density in GMCs is about $10^2$ cm\eee\ (eq. \ref{eq:barnh}).  
If GMCs collapsed in a free fall time
and formed stars, the Galactic star formation
rate would be
\beq
\dot M_*\simeq\frac{10^9~\msun}{4\times 10^6~{\rm yr}} =250~ \msun\;{\rm yr}
	^{-1},
\eeq
far greater than the observed rate in this part of the Galaxy
of 3 \sun\ yr\e\ \cite{mck97}.  This gross
disparity between the potential star formation rate
and the actual one was  pointed out many years ago by Zuckerman
and Evans \cite{zuc74}, who concluded that the
supersonic motions observed in GMCs do not reflect
gravitational infall.  Any successful theory of star formation
must account for this inefficiency.

\section{Dynamical Structure of GMCs}
\label{sec:dyng}

\subsection{THE VIRIAL THEOREM}
\label{sec:vir}

	Some insight into the structure of GMCs can be gained from
the virial theorem, as shown in the classic studies of self--gravitating
gas clouds by McCrea \cite{mcc57}
and by Mestel \& Spitzer \cite{mes56}.  Define the quantity
$I=\int r^2 dm$, which is proportional to the trace of
the inertia tensor.  Next, evaluate $\ddot I\equiv d^2I/dt^2$
from the equation of motion,
\beq
\rho\frac{d\vecv}{dt} =-{\bf\nabla}\pth+\frac{1}{4\pi} ({\bf\nabla\times
	B)\times B}+\rho\vecg,
\eeq
where $\vecg$ is the acceleration due to gravity.
The result is the virial theorem,
\beq
\frac 12 \ddot I =2(\calt-\calt_s)+\calm+\calw.
\label{eq:iddot}
\eeq
Alternatively, this relation can be derived by taking
the dot product of $\vecr$ with the equation of motion
and integrating over a volume corresponding to a fixed mass.
We shall now consider each of the terms in this
equation.

	The term on the LHS reflects variations
in the rate of change of the size and shape of the cloud.
This term is usually neglected, but it may be significant for
a turbulent cloud.  In contrast to the terms on the RHS
of the equation, it can be of either sign, and
as a result its effects can be averaged out either by
applying the virial theorem to an ensemble of clouds 
or by averaging over a time long compared with the dynamical
time.

	The first two terms on the RHS contain the
effects of kinetic energy, including thermal energy;
they do {\it not} include the energy associated with
internal degrees of freedom.
The thermal energy density inside the cloud is
$(3/2)\rho c^2=(3/2)\pth$, where $c$ is the isothermal sound speed,
and the bulk kinetic energy density 
is $(1/2)\rho v^2$.
The total kinetic energy inside the cloud is then
\beq
\calt=\int_\vcl\left(\frac 32 \pth+\frac 12 \rho v^2\right)\;
	dV\equiv \frac 32 \bar P \vcl,
\label{eq:calt}
\eeq
where $\vcl$ is the volume of the cloud
and $\bar P$ is the mean pressure of the 
gas (i.e., it does not include the pressure
associated with the magnetic field).
Note that the kinetic energy associated with
rotation is included in $\calt$ and therefore
$\bar P$; rotation is generally not a dominant
effect in molecular clouds, however \cite{goo93}
\cite{mck93}.
The mean pressure can be related to
the 1D velocity dispersion in the cloud
$\sigma$, which is observable:
\beq
\sigma^2\equiv\frac{1}{M} \int (c^2+\frac 13 v^2)\, dM
	=\frac{\bar P}{\bar\rho} .
\eeq

	The second kinetic energy term is a surface term,
\beq
\calt_s\equiv \frac 12 \int_S \pth {\bf r\cdot dS},
\eeq
where the integral is over the surface of the cloud.
If the thermal pressure in the ambient medium is constant
at $P_s$, then $\calt_s=(3/2)P_s\vcl$.  As a result,
the two kinetic energy terms combine to give
\beq
2(\calt-\calt_s)=3(\bar P-P_s)\vcl.
\label{eq:twocalt}
\eeq
Thus, it is the difference between the energy in the cloud
and that in the background medium that enters the
virial theorem.  If the ambient medium is
turbulent and
has a density much smaller than that of the cloud,
one can show that $P_s$ is somewhat
less than the total pressure 
far from the cloud \cite{mck92}:  The
stress exerted by the ambient medium normal to the cloud
is entirely thermal (and is significantly larger than the
thermal pressure far from the cloud if the turbulence
in the ambient medium is supersonic), but it is
reduced below the value it otherwise would have by
flow of intercloud gas along the cloud surface.

	The magnetic term in the virial theorem is
fairly complicated,
\beq
\calm=\frac{1}{8\pi} \int_V B^2 dV+
	\frac{1}{4\pi} \int_S {\bf r\cdot}\left(\vecB\vecB
	-\frac 12 B^2{\bf I}\right){\bf \cdot dS},
\label{eq:calm}
\eeq
where ${\bf I}$ is the unit tensor.  If the cloud is immersed
in a low density ambient medium and if
the stresses due to MHD waves in the ambient medium
are negligible, then the field outside
the cloud will be approximately force free.
In this case, one can show
that the magnetic term becomes \cite{mck92}
\beq
\calm=\frac{1}{8\pi} \int (B^2-B_0^2) dV,
\eeq
where $B_0$ is the field strength in the ambient
medium far from the cloud.  Thus,
$\calm$ is the difference between the 
total magnetic energy
with the cloud and that in the absence of the cloud.

	Finally, we consider the gravitational term $\calw$.
In the absence of an external gravitational field,
this is the gravitational energy of the cloud \cite{spi78},
\beq
\calw=\int\rho \vecr\cdot \vecg dV=
	-\frac 35 a\left(\frac{GM^2}{R} \right),
\label{eq:calw}
\eeq
where $a$ is a numerical factor of order unity that
has been evaluated by Bertoldi and McKee \cite{ber92}.
In order to relate this term to the other terms
in the virial theorem, we define the ``gravitational pressure"
$P_G$ by
\beq
\calw\equiv -3P_G\vcl;
\label{eq:calw2}
\eeq
intuitively,
$P_G$ is just the mean weight of the material in the cloud.
The gravitational pressure can be evaluated as
\beq
P_G=\left(\frac{3\pi a}{20} \right) G\Sigma^2
	\rightarrow 1.39\times 10^5\Nbhtt^2 ~~~~{\rm K\ cm^{-3}},
\label{eq:pg}
\eeq
where $\Sigma\equiv M/\pi R^2\equiv \mu_{\rm H}\Nbh$ 
is the mean surface density of the cloud.
The numerical evaluation is for a spherical
cloud with a $1/r$ density profile.

	These results enable us to express 
the steady--state, or time--averaged,
virial theorem (eq. \ref{eq:iddot} with $\ddot I=0$) as
\beq
\bar P=P_s+P_G\left(1-\frac{\calm}{\vert\calw\vert} \right).
\label{eq:pb}
\eeq
In this form, the virial theorem has an immediate intuitive
meaning: the mean pressure inside the cloud is the surface
pressure plus the weight of the material inside the cloud,
reduced by the magnetic stresses.

\subsection{ARE GMCS GRAVITATIONALLY BOUND?}
\label{sec:are}

	With these results in hand, we can now address the
issue of whether molecular clouds and their constituents
are gravitationally bound.  We assume that the cloud
is large enough that the motions are highly supersonic (\S
\ref{sec:dyn}),
and as a result the energy in internal degrees of freedom
is negligible.  The total energy is then 
$E=\calt+\calm+\calw$, which can be expressed as
\beq
E=\frac 32 \left[P_s-P_G\left(1-\frac{\calm}{\vert\calw\vert}
	\right)\right]\vcl.
\label{eq:e}
\eeq
with the virial theorem (eq. \ref{eq:pb}).
In the absence of a magnetic field, the condition that the cloud
be bound (i.e., $E<0$) is simply $P_G>P_s$.  We shall use this criterion
even for magnetized clouds, bearing in mind that using
the total ambient gas pressure 
(thermal plus turbulent) for $P_s$ is an overestimate
and that our analysis is approximate because we have 
used the time-averaged virial theorem.  In the opposite
case, in which $P_s\gg P_G$, the cloud is said to
be {\it pressure--confined}.  In the typical case
in which the pressure is largely
turbulent, the ``pressure--confined cloud" is likely to
be transient.

	For GMCs, the surface pressure is that of the ambient ISM.  
In the solar vicinity, the total interstellar pressure
is about $2.8\times 10^4$ K cm\eee,
which balances the weight of the ISM \cite{bou90}.
Of this, about $0.7\times 10^4$ K cm\eee\ is due to cosmic
rays; since they pervade both the ISM and a molecular
cloud, they do not contribute to the support of a cloud
and may be neglected.  The magnetic pressure is about
$0.3\times 10^4$ K cm\eee\ \cite{hei96}, leaving
$P_s\simeq 1.8\times 10^4$ K cm\eee\ as the total ambient gas pressure.

	What is the minimum value of $P_G$ for a molecular
cloud?  According to van Dishoeck and Black \cite{van88},
molecular clouds exposed to the local interstellar
radiation field have a layer of C$^+$ and C$^0$ corresponding
to a visual extinction of 0.7 mag (\S \ref{sec:che}).
If we require at least 1/3 of the carbon along
a line of sight through a cloud to be in the form of CO
in order to term the cloud ``molecular",
then the total visual extinction must be 
$\bar A_V>2$ mag (allowing for a shielding layer
on both sides).  According to equation (\ref{eq:pg}), this
gives $P_G\simgt 2\times 10^4$ K cm\eee\ $\sim P_s$,
verifying that molecular clouds as observed in CO 
are at least marginally bound.  Larson's ``second law"
is thus seen to be a consequence of the relationship
between the column density
required for CO to be significant and the pressure in
the ISM.
Note that if we defined molecular clouds as having
a significant fraction of H$_2$ rather than CO,
the minimum column density required would be substantially less
and the clouds might not be bound.  Furthermore,
the conclusion that CO clouds are bound depends 
on the metallicity, the interstellar pressure and the strength of the
FUV radiation field, so that CO clouds may not be bound everywhere
in the Galaxy or in other galaxies \cite{elm93a}.

	GMCs in the solar neighborhood
typically have mean extinctions significantly
greater than 2 mag, and as a result $P_G$ is
generally significantly greater than $P_s$.
Indeed, observations show  $P_G\sim 2\times 10^5$ K cm\eee,
an order of magnitude greater than $P_s$
\cite{ber92} \cite{bli91}  \cite{wil95}.
Thus, if GMCs are dynamically stable entities (the
crossing time for a GMC is about 10$^7$ yr, smaller than the expected
lifetime \cite{bli80} \cite{wil97}),
then GMCs must be self-gravitating.
In the inner galaxy, where $P_s$ is expected to be
greater, the typical GMC linewidths also appear to be somewhat
greater than those found locally \cite{san85},
and thus $P_G$ is still comfortably greater than $P_s$.

	In order to determine whether clumps within
GMCs are gravitationally bound, it is convenient
to work in terms of the {\it virial parameter} 
\cite{ber92} \cite{mye88} 
\beq
\alpha\equiv\frac{5\sigma^2 R}{GM} ,
\label{eq:alpha}
\eeq
which is readily determined from observation.  With
the aid of equations (\ref{eq:calw}) and (\ref{eq:calw2}),
this parameter can be expressed as
\beq
\alpha=a\left(\frac{2\calt}{\vert\calw\vert} \right)
	= a\left(\frac{\bar P}{P_G} \right).
\eeq
Bertoldi \& McKee \cite{ber92} show how this result
can be applied to ellipsoidal clouds; if
$\alpha$ is interpreted as an average
over the orientation of the cloud, then
the factor $a$ is generally within a factor of about
1.3 of unity.  In terms of the virial
parameter, the net energy of the cloud is
\beq
E=\vert\calw\vert\left[\frac{\alpha}{2a} -\left(1-
	\frac{\calm}{\vert\calw\vert} \right)\right].
\eeq
A clump is therefore bound for $\alpha\simlt 2$,
although the exact value depends on the strength
of the field.  For $\alpha\gg 1$,  clumps
are pressure--confined.  

	Whether a clump is pressure--confined or gravitationally
bound is directly related to its surface density \cite{ber99}.	
The surface pressure on a clump is just
the mean pressure inside the GMC, 
\beq
P_s({\rm clump})\simeq P_G({\rm GMC})\propto \Sigma^2
	({\rm GMC}),
\eeq
where we have assumed that the GMC is strongly bound,
so that the pressure acting on it can be ignored.
As a result, the
virial theorem for a clump becomes 
\beq
\bar P({\rm clump})\propto\Sigma^2({\rm GMC})+
	\Sigma^2({\rm clump}), 
\eeq
where we have assumed that the mass of an individual clump
is small, so that it does not significantly affect the surface
density of the cloud.
Pressure--confined clumps
have $\Sigma$(clump)$\ll
\Sigma$(GMC) since $P_G\ll P_s$.
On the other hand,
gravitationally bound clumps have $\Sigma$(clump)$\simgt
\Sigma$(GMC).  The studies of the stability of
gas clouds discussed below show that $P_G$ cannot be
too much greater than $P_s$ if the cloud is to
be gravitationally stable; correspondingly,
$\Sigma$(clump) cannot be much greater than
$\Sigma$(GMC) if the clump is stable.  We
conclude that gravitationally bound clumps
have column densities similar to that of
the GMC in which they are embedded,
$\Sigma$(clump)~$\sim\Sigma$(GMC),
as is often observed \cite{ber99}.

	From an analysis of \thirteenco\ clumps in
Ophiuchus, Orion B, the Rosette, and Cepheus OB3,
Bertoldi \& McKee \cite{ber92} concluded
that most clumps are pressure--confined;
however, most of the mass is in clumps that
appear to be gravitationally bound, or nearly so.
Pressure--confined clumps do not satisfy
the line width--size relation;
instead, the velocity dispersion and mean density
in the clumps are about constant. As a result,
the virial parameter scales
with mass as $M^{-2/3}$, with
the most massive clumps 
having $3\simgt\alpha\simgt 1$
(with the exception of Cepheus, for which the data
are of low resolution).  Furthermore, 
the observable star formation 
appears to be confined to these massive clumps.
Thus, not only are GMCs as a whole gravitationally bound,
but there is a mass scale $\msfc$, the mass
of the star--forming clumps, such that structures
with $M\sim\msfc$ are bound as well.

	On still a smaller scale, cloud cores (out of which
individual stars or stellar systems like binaries
form) are also observed to be gravitationally bound.
These cores exist in both star--forming clumps
and, in star--forming clouds like Taurus, in relative isolation.
Thus, there appears to be a hierarchy of bound structures:
GMCs, the star--forming clumps within them, and cloud cores
(cf \cite{fal86}).
As we shall see in \S {\ref{sec:mag2}, this hierarchy
may be mirrored in the magnetic properties of GMCs.

\subsection{ISOTHERMAL CLOUDS}
\label{sec:iso}

	Molecular gas is often observed to be at a temperature
of about 10 K.  The simplest model for a molecular cloud
is thus an isothermal cloud.  In fact, the cores of
molecular clouds often have thermal pressures that
are greater than the nonthermal pressures, so the 
model of an isothermal cloud may apply approximately
to such cores.  (For non--isothermal models, see
\cite{bol84} and \cite{fal85}.)  For now, we shall neglect the
effect of a static magnetic field.

	Using the virial theorem, we can infer the basic
properties of an isothermal, self--gravitating cloud
\cite{mcc57} \cite{spi68}.
Setting $\calm=0$ in equation (\ref{eq:pb}) and
evaluating $P_G$ with the aid of equation
(\ref{eq:pg}), we find
\beq
P_s=\frac{3M\sigma^2}{4\pi R^3} -
	\frac{3aGM^2}{20\pi R^4} ,
\label{eq:p0}
\eeq
where we have written the mean surface
density as $\Sigma=M/\pi R^2$.
If the cloud radius $R$ is large, the
second term on the RHS is negligible,
and the mean pressure in the cloud
(represented by the first term) is
approximately equal to the ambient
pressure $P_s$.  Now consider a sequence of
equilibria of smaller and smaller $R$. Initially,
reducing $R$ requires a higher
ambient pressure $P_s$.  Eventually,
the second term on the RHS, which is
$P_G$, becomes comparable to the first
and the increase in $P_s$ is halted.
This occurs when the escape velocity is
comparable to the velocity dispersion,
$GM/R\sim \sigma^2$, as can be seen by
comparing the two terms on the RHS of
equation (\ref{eq:p0}).  
Further reductions in $R$ lead to a decrease
in surface pressure, which is unstable.
The point at which $P_s$ is a maximum
therefore represents a {\it critical point}
that separates stable from unstable solutions.
Using this result for the radius
at the critical point, we find that the maximum
pressure is of order
\beq
P_{\rm cr}\propto \frac{M\sigma^2}{(GM/\sigma^2)^3}
	\propto \frac{\sigma^8}{G^3 M^2} .
\eeq
Equivalently, this relation can be interpreted as giving
the maximum mass that can be supported against
gravity in a medium of pressure $P_{\rm cr}$.
This critical mass is termed the Bonnor-Ebert
mass after the two individuals who
first worked out the structure of isothermal
spheres \cite{bon56} \cite{ebe55}:
\beqa
\mbe & = &1.18\;\frac{\dis\sigma^4}{\dis(G^3 P_s)^{1/2}} ,\\
     & = &1.15\;\frac{\dis(T/10~{\rm K})^2}{\dis
	(P_s/10^5~{\rm K~cm^{-3}})^{1/2}}
	~~~~\msun.
\label{eq:mbe}
\eeqa
The numerical value is for the conditions typical of
a low mass core, with $n\sim 10^4$~\cc\ and $T\sim 10$~K,
and it is significant that the result is of order 
the typical stellar mass.

	The maximum central density of a stable isothermal
sphere is only $\rho_c=14.0\rho_s$, where $\rho_s$ is
the density at the surface.  The mean density is
$2.5\rho_s$.  Equilibria at lower mass have significantly
lower central concentrations: for example, the
central density is only $2.2\rho_s$ for $M=0.5\mbe$.

\subsection{MAGNETIC FIELDS VS GRAVITY}
\label{sec:mag}

	Magnetic fields are believed to
play a crucial role in the 
structure and evolution of molecular clouds.
Theorists concluded that magnetic fields
are important even before molecular
clouds were discovered because simple estimates showed
that interstellar gas is far more highly magnetized
than stars are (e.g., \cite{mes56}),
and they therefore turned their attention to how this
excess flux could be lost.  
The ongoing efforts to observe magnetic fields
in molecular clouds will be discussed 
in \S \ref{sec:obsb} below.

\subsubsection{Magnetic Critical Mass}
\label{sec:mag2}

	The simplest case of a magnetically supported cloud
is one in which the field is purely
poloidal and the kinetic energy vanishes ($\calt=0$); since
the gas is cold and there are
no bulk motions, it 
settles into a thin disk.
We assume the field is fully connected with that
in the ambient medium. (The effects of closed
field lines are discussed in reference \cite{mck93}.)
For a thin disk, the volume of the cloud
vanishes, and $P_G$ goes to infinity; the virial theorem
(\ref{eq:pb}) yields
\beq
\calm=\vert\calw\vert.
\label{eq:mw}
\eeq
Let the magnetic flux threading the cloud be
\beq
\Phi=\int 2\pi r B dr\equiv \pi R^2 \bar B.
\eeq
Now consider a sequence of equilibria in which
the flux is constant and the functional form
of the mass--to--flux ratio is constant, but
the value of the mass--to--flux ratio increases.
At some point the mass--to--flux ratio will become
large enough that gravity will overwhelm
the magnetic stresses, and the cloud will collapse.
The point at which the mass--to--flux ratio 
reaches the maximum is the critical
point.  We write the net magnetic energy there as
\beq
\calm_{\rm cr}=\frac{1}{8\pi} \int (B^2-B_0^2) dV \Bigl\vert_{\rm cr}
	\equiv \left(\frac{b}{3} \right)\bar B^2 R^3
	=\left(\frac{b}{3\pi^2} \right)\frac{\Phi^2}{R} ,
\label{eq:calmcr}
\eeq
where $b$ is a numerical factor of order unity.
We can evaluate the {\it magnetic critical mass}
$\mphi$ by inserting this relation into equation
(\ref{eq:mw}) and using equation (\ref{eq:calw}):
\beq
\mphi=\left(\frac{5b}{9\pi^2 a} \right)^{1/2}\frac{\Phi}{G^{1/2}}
	\equiv c_\Phi\left(\frac{\Phi}{G^{1/2}} \right).
\label{eq:mphi}
\eeq
So long as the magnetic flux is frozen to the matter,
$\mphi$ is a constant.
For $M<\mphi$, the cloud
is said to be {\it magnetically subcritical}: 
the mass--to--flux ratio is small enough that 
magnetic stresses always exceed gravity, so such a cloud
can never undergo gravitational collapse.
Conversely,
if $M>\mphi$, the cloud is {\it magnetically supercritical},
and magnetic fields cannot prevent gravitational collapse.

	If the cloud has a constant mass--to--flux ratio
(which Shu and Li \cite{shu97} term ``isopedic"), then
the numerical factor $c_\Phi=1/2\pi$ \cite{nak78}.
Isopedic disks are highly idealized:
they can be finite
in extent only if
the field vanishes outside the disk, and
they can be in equilibrium only if they
are critical ($M=\mphi$).  
Precisely because they are so idealized, it is possible
to obtain useful results even
if they are non-axisymmetric and time-dependent \cite{shu97}.
The ratio $\mphi/M$ is constant in 
such a disk.  The distinction between
magnetically subcritical and supercritical disks
is particularly clear in this case since
the ratio of the magnetic force
to the gravitational force on a mass $M$ in the disk is (in our notation)
$(\mphi /M)^2$.

	The field strength in a cloud can be expressed in
terms of its column density by noting that
$M/\mphi\propto\Sigma/\bar B\propto \bar\Nh/\bar B$, so that
\beq
\bar B=50.5\left(\frac{\Nbhtt}{M/\mphi} \right)~~\mug
     =10.1\left(\frac{\bar A_V}{M/\mphi} \right)~~\mug.
\label{eq:barb}
\eeq

	It is sometimes convenient to have an alternative
expression for the magnetic critical mass that is
independent of the mass of the cloud.  Mouschovias
\& Spitzer \cite{mou76c} showed that such an alternative
critical mass, denoted $M_B$, is related to $\mphi$ by
\beq
\frac{M_B}{M} \equiv\left(\frac{\mphi}{M} \right)^3 .
\eeq
For an ellipsoidal cloud of size $2Z$ along the axis of
symmetry and radius $R$ normal to the axis, we have
\cite{ber92}
\beq
M_B=512\left(\frac{R}{Z} \right)^2\frac{\bar 
	B_{1.5}^3}{\bar n_{{\rm H 3}}^2}
	~~~\msun,
\label{eq:mb}
\eeq
where $\bar B_{1.5}\equiv \bar B/(10^{1.5}\;\mug)$.  This form for
the magnetic critical mass is similar in form to the Bonnor--Ebert
mass, since
\beq
M_B\propto \frac{\bar B^3}{\bar n^2} \propto\frac{1}{\bar B} \,
	\left(\frac{\bar B^4}{\bar \rho^2} \right)\propto
	\frac{v_A^4}{P_B^{1/2}} ,
\eeq
where $v_A\equiv \bar B/(4\pi \bar\rho)^{1/2}$ is the Alfv\'en velocity;
here $v_A$ plays the role of the isothermal sound speed and
the magnetic pressure $P_B\equiv
\bar B^2/8\pi$ that of the external pressure.

	Equation (\ref{eq:mb}) for the magnetic critical mass
naturally suggests three very different mass scales for
molecular clouds \cite{ber92}: (1) On the largest scales, clouds
are formed from compression of the diffuse ISM 
\cite{elm85} \cite{mes85}, which has
a mean density $\bar\nh\sim 1$~\cc\ and a field $\bar B\sim
3$~\ug; this gives $M_B\sim 5\times 10^5$ \sun,
a typical mass for a GMC (cf. eq. \ref{eq:calnc}).
Since $M_B$ is constant so long as the mass is constant
and flux freezing holds, this value will be preserved
as the gas is compressed and the GMC forms.  
(2) On intermediate
scales, the clumps within GMCs might well have originated
as the diffuse clouds in the gas that formed the GMC
\cite{elm85}.  In this case, the density is about
30--100 times greater than the average
interstellar density, but the field is about the same,
since diffuse clouds are not gravitationally bound.
As a result, we expect
$M_B\sim 50 - 500$~\sun\ in clumps, which is 
$\simlt$ the mass
of star--forming clumps [$M$(SFC)] in nearby GMCs \cite{ber92}.  
On the other
hand, $M_B$ is substantially greater than the mass
of a typical star:
Stellar mass clumps at the density of star--forming clumps
are magnetically subcritical.
(3) Finally, in regions in which the density of the
gas can grow
so that the thermal pressure is comparable to the magnetic
pressure, either by ambipolar diffusion or by flow
along field lines, then $M_B\sim \mbe\sim 1\msun$
from equation (\ref{eq:mbe}).

\subsubsection{Toroidal Fields} 

	Toroidal fields can provide a confining force,
thereby reducing the magnetic critical mass \cite{fie99} 
\cite{tom91}.  To see how
this occurs, consider a current $I(r)$ in the $z$ direction,
which generates a toroidal field $\vecB_\phi=2I/cr$.  
The resulting
force per unit volume
is $\vecJ\vectimes\vecB/c= -(J_zB_\phi/c)\hat\vecr$,
which indeed provides a pinching force.
To determine the contribution to the magnetic term in
the virial theorem, it is best to recall that the virial
theorem can be derived by taking the dot product of the 
equation of motion with $\vecr$ and integrating
over the volume; the contribution
of the toroidal field to the magnetic term is then
\beq
\calm_{\rm toroidal}=\frac{1}{c} \int (\vecJ_z\vectimes\vecB_\phi)
	\cdot\vecr dV\simeq -\frac{LI^2}{c} ,
\eeq
where $L$ is the length of the cylinder in which $\calm$
has been evaluated.  Thus, whereas poloidal fields give
a positive definite contribution to $\calm$ (provided
the field decreases outward), toroidal fields give 
a negative definite contribution.  

	Several caveats about toroidal fields should
be kept in mind:  First,
the ratio of the toroidal
field to the poloidal field cannot become too large without
engendering instabilities (e.g., \cite{jac75}).
Second, the current that generates the toroidal field
must return to where it started, since currents
do not have sources or sinks in MHD \cite{jac75}.
Once the current density reverses direction, so
does the force, and the toroidal field ceases
to be confining. If the virial theorem is
applied to a volume large enough
to encompass the entire return current,
then $I=0$ and the
toroidal field has no net effect.  In astrophysical MHD,
it is convenient to think of the current as being generated
by the field rather than vice versa; it follows
that if the toroidal field is restricted
to a finite volume, then it has no net effect on the virial
theorem applied to any larger volume.
Finally, a purely toroidal field is subject to dissipation
by magnetic reconnection, which can occur with reasonable
efficiency according to Lazarian \& Vishniac \cite{laz99}.
However, such reconnection can be avoided if there is a sufficiently
strong poloidal field along the axis, as in the protostellar wind
model of Shu et al \cite{shu94}.

\subsubsection{Clouds Supported by both Magnetic and Gas Pressure}
\label{sec:clo}

	Let us return to the case of a poloidal field, and
ask, what is the critical mass for a  cloud  supported by gas
pressure as well as magnetic stresses?  We can use the
same approach that we adopted to estimate the Bonnor--Ebert
mass \cite{mou76c}.  If we
retain the magnetic term in the virial theorem (\ref{eq:pb}),
the same steps that led to equation (\ref{eq:p0}) give
\beq
P_s=\frac{3M\sigma^2}{4\pi R^3} -
	\frac{3aGM^2}{20\pi R^4} \left(1-\frac{\calm}{\vert\calw\vert}
	\right).
\eeq
For a cloud at the critical point, $\calm=\frac 35 aG\mphi^2/R$
and $\vert\calw\vert=\frac 35 aGM^2/R$, so that $\calm/\vert\calw\vert
=(\mphi/\mcr)^2$ there.  As discussed in reference \cite{mck93},
this remains a good approximation even for subcritical clouds.
We then have
\beq
1-\frac{\calm}{\vert\calw\vert} \simeq
	1-\left(\frac{\mphi}{\mcr} \right)^2\equiv k_m. 
\label{eq:km}
\eeq
Generalizing the argument that led to the 
Bonnor--Ebert mass, we consider a sequence of equilibria
in which the ambient pressure $P_s$ is increased while
the mass and flux remain constant.  The maximum value
of $P_s$ is reached when the two terms become comparable,
which occurs at a radius such that $\sigma^2\sim GMk_m/R$.
The maximum pressure is then of order
\beq
P_{\rm cr}\propto\frac{M\sigma^2}{(GMk_m/\sigma^2)^3}
	\propto\frac{\sigma^8}{G^3k_m^3M^2} .
\eeq
For a given ambient pressure, this gives a cubic
equation for the critical mass.
If the gas is isothermal, one obtains \cite{mou76c}
\beq
\mcr=c_1\mbe\left[1-\left(\frac{\mphi}{\mcr} \right)^2\right]^{-3/2},
\label{eq:mcr}
\eeq
where $c_1$ is a numerical constant
and where we have substituted back for $k_m$ from its definition
in equation (\ref{eq:km}).  Tomisaka et al \cite{tom88}
find that their numerical results are best fit by 
$c_1=1.18$, quite close to the value estimated by Mouschovias
\& Spitzer \cite{mou76c} many years earlier.  

	An approximate solution to equation (\ref{eq:mcr}) 
for $\mcr$ is $\mcr\simeq\mbe+\mphi$,
which is accurate to within about 5\% for 
$M\simlt 8\mphi$ \cite{mck89}; for weaker fields, this approximation
predicts that $\mcr\rightarrow \mbe$, which
is presumably more accurate than the solution of equation
(\ref{eq:mcr}).  We anticipate that this
result will remain approximately valid in the more
general case in which the gas is not isothermal.
Defining the Jeans mass $\mj$ as
the critical mass associated with
thermal and nonthermal motions of the gas, we then have
\beq
\mcr\simeq \mj+\mphi.
\label{eq:mcr2}
\eeq

	Axisymmetric numerical models for magnetized clouds
were first constructed by Mouschovias \cite{mou76a} \cite{mou76b}.
He focused on the case of a mass--to--flux 
distribution corresponding to a uniform field threading
a uniform, spherical cloud.  For this case, 
Mouschovias \& Spitzer \cite{mou76c} found $c_\Phi\simeq 0.126$;
subsequent calculations by Tomisaka et al. \cite{tom88} found
$c_\Phi\simeq 0.12$, which we adopt for numerical estimates.
Tomisaka et al. showed that for a range of mass--to--flux
distributions it is the mass--to--flux ratio on the central
flux tube that controls the stability, with the critical
value being $(d\mphi/d\Phi)_c\simeq 0.17/G^{1/2}$.  
(For Mouschovias' case, they found a coefficient 
$0.18=1.5\times 0.12$;
the factor 1.5 is just the ratio of the central mass--to--flux
ratio to the mean.)  Tomisaka et al's result is quite
close to the value for an isopedic disk, $1/(2\pi G^{1/2})
\simeq 0.16/G^{1/2}$.

\subsubsection{Are Clouds Magnetically Supercritical? Theory}
\label{sec:super}

	With the framework we have developed, we can ask
the question, are GMCs magnetically supercritical or subcritical?
McKee \cite{mck89} argued that GMCs are magnetically
supercritical based on the following line of argument:
GMCs must be approximately critical $(M\simeq \mcr)$ since
they are highly pressured relative to their environment,
and calculations show that this is possible only for
nearly critical clouds \cite{mou76b} \cite{tom88}.
The large nonthermal motions observed in GMCs and the fact
that the clouds do not appear to be highly flattened
imply that $\mj$ is not small compared to $\mphi$.
Hence, from equation (\ref{eq:mcr2}), we have
$M\simeq\mcr\simeq\mj+\mphi>\mphi$.  This argument
does not set a lower bound on $\mphi$, but if
initially a cloud had $\mj\gg \mphi$ and the gas
pressure was dominated by nonthermal motions, then
the field would be amplified into approximate
equipartition, giving $\mphi\sim\mj$.  Altogether,
then, one expects GMCs to have $M\sim 2\mphi$
on theoretical grounds.  This argument was
extended to the gravitationally bound clumps
within GMCs by Bertoldi \& McKee \cite{ber92}.
These conclusions are not universally accepted,
however \cite{mou91}.

	Nakano \cite{nak98} has used similar
reasoning to conclude that the
observed cores in molecular clouds should also be
magnetically supercritical.  The general argument
that applies to all three cases---GMCs, star--forming
clumps, and cores---is that if a cloud is 
clearly gravitationally
bound, then its mass must be nearly equal to
the critical mass; if furthermore, the cloud
has more than one significant source of pressure
support, then it is supercritical with respect to
each of the sources of support individually.
Nakano then used this argument to question the
standard paradigm of low--mass star formation,
which is based on ambipolar diffusion
in magnetically subcritical clumps
\cite{mou87} \cite{shu87}.  However,
this criticism may be unwarranted: 
about half the observed
cores already contain embedded stars \cite{bei86},
so in the conventional interpretation such
cores would have already experienced substantial
ambipolar diffusion.  The issue that remains
to be resolved by observation is whether
the {\it protocores}---i.e., the precursors
of the observed cores---are magnetically subcritical.

\subsubsection{Observation of Magnetically Supercritical Clouds}
\label{sec:obsb}

	What do observations say about the strength
of magnetic fields in molecular clouds? 
The data on magnetic field strengths
come from observations of Zeeman splitting
of molecular lines, which determine $B_\parallel$,
the component of the magnetic field along the line of sight
(see Heiles et al. \cite{hei93} for a discussion
of techniques for measuring magnetic fields).
Myers \& Goodman
\cite{mye88} summarized the data available on magnetic
field strengths in 1988, and concluded that the data were
consistent with approximate equipartition among
magnetic, kinetic, and gravitational energies.
Their results are consistent with $M\simeq 2\mphi$
\cite{mck89}.  
Since in their sample the line widths vary substantially less than
the density in the clouds, the
approximate equality between kinetic and magnetic
energies ($\rho\vrms^2/2\sim B^2/8\pi$) implies
the approximate relation $B\propto\rho^{1/2}$,
as advocated by Mouschovias \cite{mou91}.
This in turn means that the \alf velocity is
independent of the density of the cloud;
Heiles et al. \cite{hei93} found that
$\langle v_A\rangle\sim 2\;\mkms$.
Subsequently, Crutcher et al. \cite{cru94} studied
the cloud B1 in some detail.  They
found that the inner envelope was marginally magnetically
subcritical, whereas the densest region was somewhat
supercritical.  The observational results were shown to
be in good agreement with a numerical model that, however,
did not include the observed nonthermal motions.

	The most comprehensive study of magnetic fields
in molecular clouds available to date is that by
Crutcher \cite{cru99}.
He concludes that molecular clouds are
generally magnetically
supercritical:  His sample, which tends
to focus on the central regions of clouds,
has no clear case
in which a cloud is magnetically subcritical.
In order to reach this conclusion, he had
to allow for projection effects:
If the magnetic field makes an angle $\theta$
with respect to the line of sight, then the
observed field $B_\parallel$ is related to the true 
field $B$ 
by $B_\parallel=B\cos\theta$, so that on average
$\langle B_\parallel\rangle=B/2$.  After allowing for this, he
finds that $\langle M/M_\Phi\rangle \simeq 2.4$
for clouds with measured fields.
Twelve of the 27 clouds in his sample have
only upper limits on the field strength;
these clouds, which typically have both lower 
densities and lower column
densities than the clouds with measured field
strengths, are also magnetically supercritical.
If the clouds are flattened along the field lines,
then the observed area is smaller than the true
area by factor $\cos\theta$ as well, so that
$M/M_\Phi\propto \cos^2\theta$; on average, this
is a factor 1/3.  However, since clouds are observed
to have substantial motions, they are unlikely
to be highly flattened along field lines,
so Crutcher concludes that $\langle M/M_\Phi\rangle \simeq 2$,
consistent with the theoretical arguments advanced
above.  Many of the upper limits
come from a study of the Zeeman effect in
OH; if these data
apply to the ``protocores" discussed
at the end of \S \ref{sec:super}, then
they indicate that the fields there
are supercritical as well, contrary to the
assumption underlying many theories of low mass star
formation. 
The clouds with detected fields are gravitationally
bound, with a mean value for the virial parameter
$\alpha\simeq 1.4$.
He also finds that the
Alfv\'en Mach number of the turbulent motions,
$m_A=\snt\surd 3/v_A$, is about unity
in the clouds with measured fields, as inferred
previously by Myers \& Goodman \cite{mye88}
on the basis of less
complete data.  Two points should be kept in
mind, however: First, these data do not address
the issue of the strength of the field on
large scales (i.e., the field threading an
entire GMC).   Second, since the data
for the detected regions
deal with dense regions in molecular clouds, it
is possible that the observed mass--to--flux ratio
has been altered by ambipolar diffusion.

\subsection{MHD WAVES IN MOLECULAR CLOUDS}
\label{sec:mhd}

	The velocity dispersions in molecular
clouds are typically $\sigma\sim 1-2$ km s\e,
whereas the thermal velocity dispersion is only
about 0.2 km s\e\ at a temperature of 10 K.
The fact that the motions in molecular clouds are
highly supersonic led Arons \& Max \cite{aro75}
to suggest that the motions in molecular clouds
are MHD waves.  This was a prescient suggestion,
made before the magnetic field measurements 
discussed above.  Subsequent discussions of
MHD waves in molecular clouds have
been given by Zweibel \& Josefatsson \cite{zwe83},
Falgarone \& Puget \cite{fal86},
Pudritz \cite{pud90}, and McKee \& Zweibel \cite{mck95}.

\subsubsection{Wave Pressure}
\label{sec:wav}

	There are three types of MHD waves:
fast, slow, and Alfv\'en.  \alf waves are particularly
simple because they have an {\it isotropic} pressure
\cite{dew70}.  At first sight, this result is surprising,
since the stress exerted by a static magnetic field
is anisotropic.  The isotropy of the pressure due
to the \alf waves can be understood as follows
\cite{mck95}: In an \alf wave the perturbation in the field
$\delta\vecB$ is orthogonal to the background field;
the stress associated with $\delta B$ gives a pressure
$\delta B^2/8\pi$ in the direction of 
the background field $\vecB_0$ and
in the direction orthogonal to both $\vecB_0$
and $\delta\vecB$.  In the direction of $\delta\vecB$,
the net stress is $-\delta B^2/8\pi$ due to
the tension in the field.  However, it is just
in this direction that the motion of the gas contributes
a dynamic pressure $\rho\delta v^2$.  Since
the kinetic and magnetic energies are in equipartition
for MHD waves \cite{zwe95}, we have
$\rho\delta v^2/2=\delta B^2/8\pi$; the net stress
along the direction of the perturbed field is then
$-\delta B^2/8\pi +\rho\delta v^2=\delta B^2/8\pi$,
which is identical to that in the other two directions.
The wave pressure is then
\beq
P_\w=\frac{\delta B^2}{8\pi} =\frac 32 \rho\snt^2,
\label{eq:pw}
\eeq
where in the last step 
we have assumed that $\delta B$ has a random orientation
with respect to the observer so that $\delta v^2=3\snt^2$.

	Understanding the dynamics of \alf waves
in molecular clouds is a complex problem in radiation
magnetohydrodynamics.  However, we can gain some
insight into the general problem by considering
simple limiting cases.  First, consider the question
of how the wave pressure varies during an adiabatic
compression \cite{mck95}.  For an adiabatic process,
we have $P_\w\propto
\rho^\gw$ for some $\gw$. The pressure is related to
the energy density $u_\w$ by 
$P_\w=(\gw-1)u_\w$.  Now, equipartition implies
\beq
u_\w=\frac 12 \rho\delta v^2+\frac{\delta B^2}{8\pi}
	=\frac{\delta B^2}{4\pi} =2P_\w,
\eeq
which in turn implies
\beq
\gw=\frac 32 .
\eeq
Thus, in a medium of uniform density $\rho(t)$,
the wave pressure varies as $P_\w(t)\propto
\rho(t)^{3/2}$.  Since $\gw$ is greater than
unity, the waves heat up during a compression
(i.e., the wave frequency increases).

	Next consider how the \alf wave pressure varies with
position in a static cloud with a density $\rho(\vecr)$.
We assume a steady state, with no sources, sinks, or losses
by transmission through the surface of the cloud.
(The latter approximation is reasonably good if there
is a large density drop at the cloud surface,
since then the transmission coefficient is small.)
We anticipate that $P_\w\propto \rho^\gpw$, where
the polytropic index $\gpw$ may differ from the adiabatic
index $\gw$.
The answer to this problem for electromagnetic radiation
is simple: the radiation pressure would be constant ($\gpw=0$).
To determine the answer for \alf waves, we
use the energy equation for \alf waves \cite{dew70} \cite{mck95}
(the $\pm$ determines the direction of propagation),
\beq
\frac{\partial u_w}{\partial t} +\vecnabla\cdot u_\w(\vecv
	\pm\vecv_A)+\frac 12 u_\w\vecnabla\cdot \vecv=0.
\eeq
In a steady state in a static cloud, this reduces to
\beq
\vecnabla\cdot u_\w\vecv_A=\vecnabla\cdot 
	\frac{u_\w\vecB}{(4\pi\rho)^{1/2}} =0,
\eeq
which implies
\beq
\vecB\cdot\vecnabla\frac{P_\w}{\rho^{1/2}} =0
\eeq
since $\vecnabla \cdot \vecB=0$ and $P_\w\propto u_\w$.
Based on this argument, McKee \& Zweibel \cite{mck95}
concluded that
$P_\w(\vecr)\propto\rho(\vecr)^{1/2}$
along any field line.  If the constant of
proportionality is the same for all the field
lines, then the \alf wave pressure satisfies
a polytropic relation $P_\w(\vecr)\propto \rho(\vecr)^\gpw$ 
with 
\beq
\gpw=1/2.
\eeq
This result is consistent with
that of Fatuzzo \& Adams \cite{fat93}, who studied
the particular case of \alf waves in a self-gravitating
slab threaded by a uniform vertical magnetic field.
The fact that $\gpw$ is less than unity means that
the velocity dispersion increases as the density decreases:
the surface of the cloud is ``hotter" than the center,
which is consistent with the observed line width--size
relation.  The fact that the polytropic index
$\gp$ differs from the adiabatic index $\gamma$
introduces a complication into modeling molecular
clouds, as we shall see below.

\subsubsection{Wave Damping}
\label{sec:wav2}

	MHD waves are subject to both linear and nonlinear
damping (see \S \ref{sec:mod} for the latter).  
The linear damping is due to ion--neutral
friction, the same process that governs ambipolar
diffusion.  At sufficiently low frequencies, the
ions and neutrals are well coupled so that the
\alf velocity is determined by the density of
the entire medium, $\rho=\rho_n+\rho_i$.  
For transverse waves the equation of motion
for the neutrals is
\beq
\rho_n\frac{d\vecv_n}{dt} = \rho_n\nuni \vecv_{D},
\eeq
where $\nuni$ is the neutral--ion collision frequency and
$\vecv_D$ is the velocity of the ions with
respect to the neutrals.  For a linear wave of
frequency $\omega$ and velocity amplitude
$\delta v$, this yields 
$v_D=\omega\delta v/\nuni$.  The specific energy 
of the waves (including both kinetic and magnetic
energy) is $\epsilon=\delta v^2$; the rate
at which this energy is damped out is
$\dot\epsilon=\nuni v_D^2$.  In terms of the 
damping rate for the wave amplitude $\Gamma$, we
have $\dot\epsilon=-2\Gamma\epsilon$, so that
\beq
\Gamma=\frac{\omega^2}{2\nuni} .
\eeq
This heuristic argument, which implicitly assumes
$\Gamma\ll\omega$, suggests that
the waves are critically damped
(i.e., $\Gamma=\omega$) at a frequency $\omega_{\rm cut}=2\nuni$,
corresponding to a wavenumber $\kcut=\omega_{\rm cut}/v_A=
2\nuni/v_A$.  A more precise calculation \cite{kul69} shows
that the real part of the frequency vanishes at 
$k=\kcut$.  MHD waves in which the motion of the ions
and neutrals is coupled cannot propagate if
$k>\kcut$, and they therefore 
cannot provide pressure support on scales 
smaller than about $R_{\rm cut}\equiv\pi/\kcut$.

	The numerical value of $R_{\rm cut}$ depends
on the neutral--ion collision frequency,
$\nuni=n_i\langle \sigma v\rangle=1.5\times 10^{-9}n_i$~s\e\
\cite{nak84}.  Using equation (\ref{eq:xe}) for the ionization,
we find that for a cloud of radius $R$ and mass--to--flux
ratio governed by $M/\mphi$,
\beq
\frac{R}{R_{\rm cut}} =7.7\left(\frac{C_i}{10^{-5}\;\mcm^{-3/2}}
	\right)\frac{M}{\mphi} .
\eeq
Thus, for typical levels of ionization produced by cosmic
rays (\S \ref{sec:che}), 
clouds large enough that they cannot be supported by static
magnetic fields alone ($M>\mphi$) are large enough to 
support a modest spectrum of MHD waves ($R\gg R_{\rm cut}$)
\cite{mck93} \cite{wil98}.

\subsection{POLYTROPIC MODELS FOR MOLECULAR CLOUDS}

	Polytropic models, in which the pressure
varies as a power of the density,
\beq
P=\kp\rho^\gp,
\eeq
have long been used to model stars.  
The power $\gp$ is often expressed in terms of
an index $n$,
\beq
\gp\equiv 1+\frac{1}{n} .
\eeq
Prior to
the advent of computers, polytropic models were the best 
available for studying stellar structure,
and even now they are useful for gaining
insight.  Models of
molecular clouds are decades behind those for stars,
with computational models only now beginning to be
developed (\S \ref{sec:mod}).  Thus, polytropic models can
be expected to be of use here too, in order to shed
light on the density structure of molecular clouds,
the line width---size relation, the precollapse conditions
for star formation, and the relation between the properties
of GMCs and the medium in which they are embedded.

	The isothermal Bonnor--Ebert models discussed 
above are the simplest examples of polytropes.
Non-isothermal polytropes ($\gp\neq 1$) have been
discussed by Shu et al. \cite{shu72}, Viala \& Horedt
\cite{via74}, and Chi\'eze \cite{chi87}.  Maloney
\cite{mal88} pointed out that the line width---size
relation demanded a ``negative--index" polytrope,
in which $\gp<1$ so that $n$ is negative.
In order to treat the nonthermal motions in molecular
clouds, Lizano \& Shu \cite{liz89} 
assumed that the pressure associated with
these motions is proportional to the 
logarithm of the density---the limiting
case of a negative index polytrope in which
$\gp\rightarrow 0$.  This 
``logatropic" form for the 
turbulent pressure gives a sound speed
$(dP/d\rho)^{1/2}\propto 1/\rho^{1/2}$, which is consistent with
Larson's laws (\S \ref{sec:dyn}); on the other 
hand, it as yet has no physical basis.  The
logatropic equation of state has been studied
further by Gehman et al. \cite{geh96} and, in a different
form, by McLaughlin \& Pudritz \cite{mcl96}.

\subsubsection{Structure of Polytropes}
\label{sec:str}

	The structure of a polytrope is controlled by the
value of $\gp$.  Some insight into the behavior of 
spherical polytropes
can be gained by considering the limiting case of
singular polytropic spheres,  which have power law
solutions.  From the equation of hydrostatic
equilibrium
\beq
\frac{dP}{dr} =-\frac{GM\rho}{r^2} ,
\eeq
one readily finds \cite{cha39}
\beq
\rho  \propto  r^{-2/(2-\gp)},~~ 
P    \propto  r^{-2\gp/(2-\gp)},~~ 
c     \propto  r^{(1-\gp)/(2-\gp)},
\eeq
where $c^2=P/\rho$ is the generalized isothermal sound speed.
A singular isothermal sphere \cite{shu77}, for example,
has $\gp=1$ so that $\rho\propto P\propto 1/r^2$
and $c=$const.  For a cloud supported by \alf waves,
we have $\gp=1/2$ so that $\rho\propto r^{-4/3}$ and
$c\propto r^{1/3}$.  In the limit as
$\gp\rightarrow 0$, which is an approximation for a logatrope,
we have $\rho\propto 1/r$ and $c\propto r^{1/2}$.
Note that in the last two cases the velocity
dispersion increases outward,
as observed.  However, these simple power law
models cannot be used to determine the nature
of the pressure support in molecular clouds,
both because actual polytropes are not power laws
and because actual clouds have more than one
source of support.

	Since the Bonnor--Ebert mass scales as
$c^3/(G^3\rho^{1/2})$, it is convenient to
introduce a dimensionless mass  \cite{mck99} \cite{sta83}
\beq
\mu\equiv \frac{M}{c^3/(G^3\rho)^{1/2}} =
	\frac{M}{c^4/(G^3 P)^{1/2}} .
\label{eq:mu}
\eeq
For stars, which have $\gp>4/3$, this quantity can go
to infinity at the surface: stars are supported by
the hot gas in their interiors.  However, the sources
of support for molecular clouds have $\gp\leq 4/3$,
and for such clouds there is an upper limit on
$\mu$ of 4.555, so that stable clouds satisfy \cite{mck99}
\beq
M< 4.555\left(\frac{c_s^4}{G^{3/2}P_s^{1/2}} \right) ,
\label{eq:mless}
\eeq
where the subscript ``s" emphasizes
that the sound speed and pressure are evaluated at the surface
of the cloud.  Thus,
the mass of a molecular cloud is limited by conditions
at its {\it surface}.  By contrast, the maximum mass
of a star is set by conditions at or near its center.
This upper limit on $\mu$ is
a monotonically increasing function of $\gp$; for
negative index polytropes
($\gp<1$), it must be less than the Bonnor--Ebert value, $\mu=1.18$.

\subsubsection{Stability of Polytropes: Locally Adiabatic Components}
\label{sec:stab}

	In order to determine the stability of a cloud, we need
to know how it will respond to a perturbation.
We shall assume that the perturbation can be modeled
as being adiabatic, in the sense that there is
no heat exchange between pressure components
(e.g., we ignore wave damping)
and there is no heat exchange with the environment
(e.g., we ignore the loss of wave energy by transmission
into the ambient medium).
McKee \& Holliman \cite{mck99} distinguish two
cases: if the heat flow associated with the given
pressure component is very inefficient, then
the component is {\it locally adiabatic}.
In the opposite limit of very efficient heat flow,
the component is
{\it globally adiabatic}.
An adiabatic gas in conventional parlance is
locally adiabatic.  
For a locally adiabatic component, the 
entropy parameter
\beq
K_i\equiv\frac{P_i}{\rho^\gi}
\eeq
remains constant during the perturbation,
where the adiabatic index $\gi$ is distinct
from the polytropic index $\gpi$.
(The actual entropy is proportional to the 
logarithm of the entropy parameter.)
The magnetic field can be modeled approximately as
a locally adiabatic component with $\gamma_B=4/3$;
the corresponding entropy parameter is $K_B\propto
B^2/\rho^{4/3}\propto M_B^{2/3}$, which
indeed is constant so long as flux freezing holds.

	If the adiabatic and polytropic indexes are
the same ($\gi=\gpi$), then the gas is
{\it isentropic} since the entropy parameter 
$K_i\propto \rho^{\gpi-\gi}$
is spatially and temporally constant.
Even if the gas is subject to heating and
cooling, it may be possible to model it as
an isentropic gas:
If the heating rate scales as
$nT^a$ and the cooling rate as $n^2T^b$, 
where $a$ and $b$ are constant, then 
the gas can be modeled as isentropic with 
$\gi=\gpi=1+1/(a-b)$.  A discussion of
the value of $\gi$ based on molecular cooling
curves and allowing for variable $b$
is given by Scalo et al. \cite{sca98}.
Note that if the gas is not isentropic,
then perturbation of a polytrope leads to
a configuration that is not polytropic.

	A polytrope supported by a locally adiabatic
pressure component is stable for
$\gi>4/3$.  For $\gi<\gpi$, the polytrope is
convectively unstable.  Along the line
$\gi=\gpi$ (isentropic polytropes), the
critical point that divides unstable 
clouds from stable ones lies at the 
maximum value of $\mu$; equation
(\ref{eq:mless}) thus gives an upper limit on
$\mcr$.  As $\gi$
increases above $\gpi$, the value of $\mu$ at the critical point
($\mucr$) changes, but it remains close to, and somewhat
less than, the maximum value of $\mu$.
The magnitude of the density contrast
between the center of the cloud and the surface,
$\rhoc/\rhos$, increases dramatically
as $\gi$ increases, and can become infinite
even if $\gi$ is less than 4/3 \cite{mck99}.

	For isothermal polytropes, $\mcr$ is reduced
somewhat by an increase in the external pressure, $\mcr\propto
 P_s^{-1/2}$ (eq. \ref{eq:mless}).  
For negative--index polytropes, however,
$\mcr$ can decrease much more sharply with $P_s$
due to the decrease in $c_s$ \cite{shu72}.
Since the decrease in the temperature is bounded
(it is difficult to cool below 10 K in a typical
molecular cloud, for example), it is more convenient
to express the critical mass in terms of quantities
after the compression,
\beq
\mcr=\mucr\left(\frac{c_{s,f}^4}{G^{3/2} P_{s,f}^{1/2}} \right) ,
\label{eq:mcr3}
\eeq
where $c_{s,f}$ is the final value of $c_s$, etc.
This result shows that the 
reduction in the critical mass due to cooling,
which reduces $c_s$, can be large, but the reduction due to
the compression is limited by the weak $P_{s,f}^{-1/2}$
dependence.  For example, 
in a radiative shock the final pressure 
is related to the initial value $P_{s,i}$ by
$P_{s,f}=(v_{\rm shock}/c_{s,i})^2 P_{s,i}$.
In spherical implosions even higher compressions,
and correspondingly greater reductions in $\mcr$,
are possible \cite{toh87}, although in practice
it may be difficult to maintain the high degree
of spherical symmetry required to achieve very 
large compressions.

\subsubsection{Stability of Polytropes: Globally Adiabatic Components}
\label{sec:stab2}

	MHD waves are not locally adiabatic since they
can move in response to a perturbation.  In the absence
of damping and losses, or in the case in which sources balance
damping and losses, it is the wave action integrated over the cloud
that is conserved
\cite{dew70}, and the waves can be said
to be ``globally adiabatic."  The wave action is related to
the entropy parameter for the waves \cite{mck95},
and McKee \& Holliman \cite{mck99} have determined
how to treat the stability of a globally adiabatic pressure
component.  This is a generalization of the problem
of the stabiity of globular clusters considered by
Lynden--Bell \& Wood \cite{lyn68}, with the complication
that the gas is not isothermal.  Lynden--Bell \& Wood modeled
globular clusters as polytropes with $\gp=1$
and $\gamma=5/3$.  Whereas locally adiabatic polytropes
are stable for $\gamma>4/3$, this is not the case
for globally adiabatic polytropes.  If the density
contrast between the center and edge becomes too
great, the cluster is subject
to core collapse, in which the stars carry heat from the core to
the envelope and  allow the core to collapse while
the envelope expands.  This is a generic property
of globally adiabatic polytropes: for $\gp<6/5$,
such polytropes are unstable for arbitrary values
of $\gamma$, with global collapse occurring for
$\gamma<4/3$ and core collapse for $\gamma>4/3$.
Since $\gw=3/2$, a polytrope supported by \alf waves
is subject to core collapse.  Note that for
$\gamma>4/3$, the cloud heats up and becomes
more stable if it is compressed; it is {\it decompression}
that leads to instability.

	Because $\gpw$ is only 0.5, the critical mass
for a cloud supported by \alf waves is smaller
than that for an isothermal cloud.  Calculations
show that the critical mass for such a cloud is \cite{mck99}
\beq
\mw=0.39\left(\frac{\sigma_{{\rm nt},s}^3}{G^{3/2}\rhos^{1/2}}
	\right)=0.65\left(\frac{\langle\snt^2\rangle^
	{3/2}}{G^{3/2}\rhos^{1/2}} \right).
\eeq
A cloud supported by the pressure of both an isothermal
gas and \alf waves has a critical mass \cite{mck99}
\beq
\mj=1.18\left(\frac{\sigma_{\rm eff}^3}{G^{3/2}\rhos^{1/2}} \right),
\eeq
where the effective velocity dispersion is
\beq
\sigma_{\rm eff}^2\equiv c_{\rm th}^2 + 0.67\langle\snt^2\rangle.
\eeq
Insofar as it is accurate to represent the nonthermal
motions in clouds as \alf waves, they are less effective
at supporting clouds than thermal motions
because they tend to concentrate in the low
density envelope, as indicated by their polytropic
index $\gpw=1/2$.

\subsubsection{Multi--Pressure and Composite Polytropes}
\label{sec:mul}

	Real clouds are supported by thermal pressure,
magnetic stresses, and wave pressure.
Polytropic models with multiple components 
can be classified into
three types \cite{mck99}: {\it Composite polytropes}, in which
the pressure components are spatially separated
(an example is the core--envelope model for
red giant stars of Sch\"onberg \& Chandrasekhar
\cite{sch42}); {\it multi--fluid polytropes}, in which
the different components interact only gravitationally,
as in the case of a molecular cloud with an embedded
star cluster;
and {\it multi--pressure polytropes},
in which there is a single self--gravitating fluid
with several pressure components,
\beq
P(r)=\Sigma P_i(r)=\Sigma \kpi \rho^\gpi.
\eeq

	Lizano \& Shu \cite{liz89} developed 
the first multi--pressure polytropic model for molecular clouds.
They treated the axisymmetric magnetic field exactly,
and modeled the gas pressure as consisting of an
isothermal component for the thermal
pressure and a logatropic component
for the turbulent pressure.
Gehman et al. \cite{geh96} studied the
stability of logatropes in both planar
and cylindrical geometries; they effectively
assumed an isentropic equation of state, which
does not allow for either the stiffness or the mobility of
the \alf waves.  McLaughlin
\& Pudritz \cite{mcl96} studied a variant
of the logatrope in spherical geometry.
To assess the stability of the 
clouds, they adopted the boundary condition 
proposed by Maloney \cite{mal88} in which
the central temperature is held constant.
While this is plausible for the thermal pressure,
no justification has been advanced for using
it for the wave pressure.

	An approximate alternative to the multi--pressure
polytrope has been developed by Myers \& Fuller \cite{mye92}
and Caselli \& Myers \cite{cas95}.  In the ``TNT" model,
the density is assumed to obey 
\beq
n\propto \left(\frac{r_0}{r} \right)^2+\left(\frac{r_0}{r}
	\right)^p,
\eeq
where the two terms represent the effect of thermal and
nonthermal motions, respectively.  This form for the density
is inserted into the equation of hydrostatic equilibrium
to determine the density at $r_0$ and the velocity dispersion
as a function of $r$.  In most cases it is possible to
obtain a good fit to data on the line width as a function 
of $r$ by fitting the characteristic length scale
$r_0$ and the exponent $p$.  Caselli \& Myers find that
``massive cores"
have significantly smaller values of $r_0$
(0.01 pc vs. 0.3 pc) and larger mean extinctions
($A_V=15$ mag vs. 3.6 mag) than ``low mass cores"
(although the values of the masses are not discussed).

	A preliminary study of multi--pressure polytropes
including thermal pressure, magnetic pressure,
and \alf waves
has been given in John Holliman's thesis \cite{hol95}.
An important result from this work is that when allowance
is made for the layer of gas in which the C is atomic,
which has a thickness
of 0.7 mag \cite{van88}, stable clouds can have a mean gas pressure
in the CO up to about 8 times the gas pressure acting
on the surface of the cloud. (Elmegreen
\cite{elm89} had previously considered the pressure
due to the smaller layer of HI, and did not
address the stability of his model.) Applying the virial
theorem (eq. \ref{eq:pb}) to the CO, we have
$\bar P({\rm CO})\simeq P_G({\rm CO})$.  Since
$P_G({\rm CO})\simeq  1.4\times 10^5 \Nbhtt^2({\rm CO})$ K cm\eee\
from equation (\ref{eq:pg}), the mean column density
associated with the CO is
\beq
\Nbhtt({\rm CO})\simeq 1.0\left(\frac{P_s}{2\times 10^4\;{\rm K\;cm^
	{-3}}} \right)^{1/2}.
\eeq
Allowing for the somewhat higher ambient pressure 
in the inner Galaxy and in regions of
active star formation, and for the atomic
gas associated with the GMCs,  we infer
\beq
0.4\simlt\Nbhtt\simlt 2,
\eeq
where the lower limit is set by the condition that
there be a significant column density of CO
(\S \ref{sec:che}).  This argument, which builds
on previous work by Chi\'eze \cite{chi87}
and Elmegreen \cite{elm89}, provides an explanation
for Larson's third law:  All molecular clouds have about
the same column density since if the column is too
low they are not molecular and if it is too high
they are not stable.  The argument near the end
of \S \ref{sec:are} shows why Larson's law then
applies to typical star--forming clumps in GMCs as well. 
OB star--forming clumps have far larger extinctions
($\Nbhtt\sim 60$) \cite{plu97}, 
and so do not satisfy Larson's third law; their stability
needs investigation.

	As discussed in \S \ref{sec:dyn}, any one of 
Larson's laws can be derived from the other two.
To obtain the line width---size relation, we write
the velocity dispersion in the cloud 
in terms of the virial parameter $\alpha$ (eq. \ref{eq:alpha}),
\beq
\sigma=0.55\left(\alpha\Nbhtt R_{\rm pc}\right)^{1/2}~~~~\mkms.
\eeq
For clouds that are gravitationally bound ($\alpha\sim 1$), the 
fact that the mean column density is restricted to a fairly
narrow range of values then leads to the observed
line width---size relation, $\sigma\propto
R^{1/2}$.  The line width also depends
on the magnetic field strength:
Since $\bar A_V\propto \bar B/(M/\mphi)$
from equation (\ref{eq:barb}), we see that
$\sigma\propto \bar B^{1/2}$ \cite{mou95} \cite{mye88}
provided $M/\mphi$ is about the same for all the clouds
as indicated in \S\S \ref{sec:super} and \ref{sec:obsb}.

	A limitation of the polytropic models described above is
that they do not allow for the damping of MHD waves
on small scales.  Curry \& McKee
\cite{cur99} have developed
composite polytropic models to deal with this problem.
A simple example of a composite polytrope is
one in which a cold isothermal core is embedded
in a hot isothermal envelope; in this case,
it is possible to have a total pressure drop
between the center of the cold core and the surface
of the hot envelope of up to $14^2=196$.  To
model molecular cloud cores, they assume that
the inner region 
is dominated by thermal motions whereas the outer region is
dominated by nonthermal motions and a static field.
By carefully allowing for projection effects, they
are able to get good agreement with the observed
line width profiles.

\subsection{MODELING TURBULENCE IN MOLECULAR CLOUDS}
\label{sec:mod}

	The most important recent development in the
theoretical study of molecular clouds is the advent
of sophisticated simulations of turbulent, magnetic,
self--gravitating clouds \cite{gam96}, \cite{mac98a}, 
\cite{mac98b}, \cite{ost98}, \cite{pad97}, \cite{pad98},
\cite{sto98}.
Many of the issues associated with turbulence
in molecular clouds and the results from these
simulations have been reviewed by Vazquez--Semadeni
et al. \cite{vaz99}.  Here I shall mention only
two results that are of particular relevance to our
discussion.

	First, the simulations show that when the 
turbulent velocities are supersonic, the gas becomes
clumped.  As yet, there is no agreement 
on exactly how the clumping factor depends on
the physical conditions.   Vazquez--Semadeni et al
\cite{vaz99} summarize the existing 
multi--dimensional, isothermal calculations as finding
that the mean mass--averaged value of 
$\log(\rho/\bar\rho)$ scales with the logarithm
of the sonic Mach number.
Since the Mach number increases with scale,
this result suggests that the clumping factor
in clumps is smaller than in GMCs as a whole.
To date, there is insufficient dynamical range
in space to study hierarchical structure in GMCs
(cores in star--forming clumps in GMCs),
or in time to study how the initial conditions
in the interstellar medium might affect the
clump structure of GMCs.

	The second and more important result is that
the waves damp remarkably quickly \cite{mac98a}, \cite{mac98b},
\cite{sto98}.  For example,
Stone et al. \cite{sto98} find that the
energy dissipation time for forced MHD turbulence
is about $0.75L/\vrms$, where $L$ is the scale on which
the turbulence is supplied and $\vrms=\surd 3\snt$ is the rms velocity
of the turbulence.  None of the groups find a significant
difference between magnetized and unmagnetized turbulence.
This is completely contrary to the expectation
that magnetic fields would reduce dissipation
by ``cushioning" the flow.  Furthermore,
since circularly polarized \alf waves of arbitrary
amplitude can propagate without dissipation in a uniform
medium, it had been thought that they could survive
more than an eddy turnover time in a non--uniform medium.
Stone et al estimate that the energy dissipation rate per
unit mass is $\dot\epsilon_{\rm diss}\simeq \vrms^3/L$.
The motions observed in molecular clouds must therefore
be continually rejuvenated, presumably by energy injection
from newly formed stars \cite{nor80}.

	Although several
groups have independently found that the waves
damp extremely rapidly, this result must be treated with caution.
From a technical standpoint, the numerical resolution is as yet
inadequate to resolve the waves that occur in the clumps
that form within the simulation volume; this could be
important since such waves could resist compression of
the clumps.  To date, simulations have been done
for gas that fills a rectangular volume; it has not been possible to 
simulate an isolated GMC.
More importantly, there are several
problems from an observational standpoint.
It is very difficult to understand how a GMC such
as G216-2.5, which has no visible star formation, 
can have a level of turbulence that
exceeds that in the Rosette molecular cloud,
which has an embedded OB association \cite{wil99}.  
On a smaller scale, cores are observed to have
comparable levels of non-thermal motions whether
there is an embedded star or not (e.g., \cite{cas95}).
One of the striking results of the simulations
is that the rate of dissipation appears to
be insensitive to whether the turbulence is
super--Alfv\'enic or not, yet observations
show that the \alf Mach number is of order
unity \cite{cru99}.  Although there is
considerable scatter in the properties of molecular
clouds, the existence of regularities such as those
found by Larson \cite{lar81} is difficult to understand
if the turbulence decays on the dynamical time scale
whereas the stars that are supposed to support the clouds
form on a considerably slower time scale, as discussed
in \S \ref{sec:star}.

\section{Evolution of Molecular Clouds and Star Formation}

\subsection{FORMATION OF GMCS}

	The formation of GMCs is a rich and complex topic
that has been reviewed by Elmegreen \cite{elm93b}.
As in the case of galaxy formation, there are two countervailing
views: in the ``bottom--up" picture, small objects coagulate
to form large ones, whereas in the ``top--down" picture,
large objects form first and fragment into small objects.
In the case of GMCs, the bottom--up scenario 
corresponds to the growth of clouds by collisions.
In order to generate the observed power--law mass distribution,
a number of generations of collisions are necessary.
However, as Elmegreen \cite{elm93b} points out, the short
lifetime of molecular clouds (\cite{bli80} \cite{wil97})
makes this very difficult.  

	In the top--down scenario of GMC formation,
clouds form in spiral arms, where the low shear
and high densities allow gas to accumulate along
the arm \cite{elm94}.  The volume from which the
gas in a GMC is accumulated is quite large \cite{mes85}.
If we assume that the accumulation volume 
prior to compression by the spiral arm is an 
ellipsoid with a axial diameter $L_0$ and a radius $R_0$,
that the mean density in this volume is initially
$n_0$, and that the magnetic field (measured in \ug ) is
initially $\bom$, then one finds \cite{mck93}
\beqa
L_0 & = & 96\left(\frac{\bom}{n_0} \right)\frac{M}{\mphi} 
	~~~~{\rm pc},\\
R_0 & = & 380 \left(\frac{M_6}{\bom} \right)^{1/2}
	\left(\frac{M}{\mphi} \right)^{-1/2} ~~~~{\rm pc}.
\eeqa
For typical conditions ($n_0\sim 1$ cm\eee, $\bom\sim 3$
and $M\simeq 2\mphi$, as discussed in \S \ref{sec:super}), this gives
$L_0\sim 600$ pc and $R_0\sim 150 M_6^{1/2}$
pc.  This is large enough to contain many diffuse clouds,
which could subsequently become clumps in the GMC
\cite{elm85}.

\subsection{DYNAMICAL EVOLUTION OF GMCS}

	Once a GMC has formed, it will be supported against
gravity by a combination of static magnetic fields and
turbulent pressure; since the observed motions are
highly supersonic, thermal pressure is relatively unimportant.
Because the waves are damped (\S \ref{sec:mod}), the
cloud will contract.  The contraction of the cloud
will adiabatically compress the waves, tending to 
counteract the damping.  In order to stop the contraction,
energy injection is required, and this is provided by
protostellar outflows \cite{nor80}.

	The resulting evolution of the cloud can
be described with the ``cloud energy equation" \cite{mck89}.
Let $\epsilon\equiv E/M$ be the total cloud energy per unit mass.
Since the motions observed in GMCs are highly supersonic,
we shall assume that the energy in internal degrees
of freedom is negligible, as in \S \ref{sec:are}.
The rate of change of $\epsilon$ can be written
symbolically as
\beq
\frac{d\epsilon}{dt} =\calg-\call,
\label{eq:eps}
\eeq
where $\calg$ is the rate of energy gains per unit mass and
$\call$ is the rate of energy losses per unit mass.

	Energy is lost by wave damping, $\call=-\dot E_w/M$,
where the wave energy $E_w$ includes the energy in both
motions and in fluctuating fields.  For motions coupled
to the field, the fluctuating field energy is in equipartition with
the kinetic energy \cite{zwe95}.  If we assume that the motions
are isotropic, then equipartition applies
to 2 of the 3 directions and $E_w=(5/3) \calt$.
In fact, Stone et al. \cite{sto98} find $E_w\simeq 1.6 \calt$
when the gas is strongly magnetized ($\beta\equiv \pth/P_B$
in the range 0.01-0.1).  Let $\eta$ be the ratio of the
damping time to the free fall time $\tff$;
we then have
\beq
\call\simeq 1.6\,\frac{\calt/M}{\eta\tff} =1.6\,\frac{
	\vrms^2/2}{\eta\tff}.
\label{eq:call}
\eeq
The simulations discussed in \S \ref{sec:mod} suggest
that $\eta\sim 1$, but for the reasons discussed there
$\eta\sim$~a~few seems more reasonable.

	As pointed out by Norman \& Silk \cite{nor80},
energy injection by newly formed stars is
the dominant source of kinetic energy for molecular
clouds.  The rate at which protostellar outflows inject energy 
can be written as
\beq
\calg=\epsin\left(\frac{\dot M_*}{M} \right)\equiv
	\frac{\epsin}{\tgs} ,
\label{eq:calg}
\eeq
where $\epsin$ is the energy injected per unit stellar mass
and $\tgs$ is the {\it star formation time scale}, 
defined as the time
to convert the gas entirely into stars.
An outflow drives a shock into the surrounding medium and
a reverse shock is driven back into the outflow.  Under
the assumption that both shocks are radiative, 
momentum conservation implies $m_wv_w=
M_{sw}v_{sw}$, where $m_w$ and $v_w$ are the mass and velocity
of the outflowing wind, and $M_{sw}$ and $v_{sw}$
are the mass and velocity of the swept up cloud.  
The swept up material merges with the ambient cloud
when $v_{sw}$ drops to about the effective
sound speed $(P_{\rm tot}/\rho)^{1/2}$, which
is of order $\vrms$ in a highly turbulent cloud.
Introducing a factor $\phi_w\sim 1$ that allows for the
uncertainty in our estimate of the energy injection
per outflow, we then have
\beq
m_*\epsin = 1.6\phi_w\times \frac 12 M_{sw}\vrms^2=
	1.6\left(\frac{\phi_w m_wv_w}{\vrms} \right)\frac{\vrms^2}{2} ,
\eeq
where we have included a factor 1.6 for the energy stored
in fluctuating magnetic fields.  Magnetic energy stored in the 
wind could do work on the ambient medium, which
would make $\phi_w$ greater than unity; on the 
other hand, the protostellar jets could escape from
the cloud, which would reduce $\phi_w$.
Inserting this result
into equation (\ref{eq:calg}) and using equations
(\ref{eq:eps}) and (\ref{eq:call}), we find that the
cloud energy equation becomes
\beq
\frac{d\epsilon}{dt} =1.6\left[\left(\frac{\phi_wm_wv_w}{m_*\vrms} 
	\right)\frac{1}{\tgs} -\frac{1}{\eta\tff} \right]
	\frac{\vrms^2}{2} .
\eeq

	This equation shows that the cloud will contract if the
star formation rate is too small 
(for a bound cloud, $\epsilon$ will become more negative)
and expand if the rate is too large.  
If the cloud contracts quasistatically (so that $\ddot I$
is negligible), and if the cloud is strongly bound 
($P_s\ll\bar P$, which may be difficult to achieve in
practice), then the virial
theorem implies $E=\calt+\calm+\calw=-\calt$,
or $\epsilon=-\vrms^2/2$.  Thus, contraction of 
a bound cloud leads to an {\it increase} in the velocity
dispersion of the cloud.  This ``virialization" might
account for the motions observed in clouds that are
not actively forming stars \cite{mck89}.

    As the density of the contracting cloud rises,
the star formation rate is expected to rise as well.
If the rate of energy injection by newly formed
stars balances the damping rate, the star formation
is said to be {\it dynamically regulated}
\cite{ber99}:
\beq
\frac{\tgsdr}{\tff} =\left(\frac{m_wv_w}{m_*\vrms} \right)\phi_w\eta.
\eeq
Theoretically, we expect the momentum per unit
stellar mass to be 
\beq
\left(\frac{m_w}{m_*} \right)v_w 
		      \simeq \frac 13 \times 200\,\mkms \simeq 70\,\mkms
\eeq
\cite{naj94}.  Observations suggest a
somewhat lower value of about 40 \kms\ \cite{ber99}.
With the latter value, and taking $\vrms\sim 2$ \kms, we have
$\tgsdr/\tff\simeq 20 \phi_w\eta\gg 1$.  Thus,
dynamically regulated star formation is inefficient,
consistent with observation.

	   To obtain a more precise comparison 
between the dynamically regulated star
formation rate and observation, we write the 
dynamically regulated star
formation rate as
\beq
\dot M_{*\dr}=\frac{M}{\tgsdr} =\left(\frac{m_*\vrms}{m_wv_w} \right)
	\frac{M}{\phi_w\eta\tff} .
\eeq
Noting that $\vrms/\tff\propto \alpha^{1/2} \Nbh$
and using 40 \kms\ for the momentum injection per unit stellar mass,
we find
\beq
\dot M_{*\dr}\simeq 2.7\times 10^{-8}\left(\frac{\alpha^{1/2}\Nbhtt 
	M}{\phi_w \eta} \right)~~~~\msun~{\rm yr}^{-1}.
\eeq

	First, apply this equation to the entire Galaxy.
The mass of $\htwo$ inside the solar circle
is $1.0\times 10^9$ \sun\ \cite{wil97}, of which
about half is actively forming stars \cite{sol89}.
With an average column density $\Nbhtt\simeq 1.5$
\cite{sol87} and with $\alpha\simeq 1$ \cite{ber92},
we have $\dot M_*\simeq 20/(\phi_w\eta)$ \sun\ yr\e.
This is consistent with the observed value of
3 \sun\ yr\e\ \cite{mck97} for $\phi_w\eta\simeq 7$.
McKee \cite{mck89} adopted a momentum per unit stellar
mass of 70 \kms\ instead of 40 \kms (which is equivalent
to $\phi_w=1.7$) and took $\eta=5$, and so concluded
that energy injection by low mass stars could supply the
turbulent energy needed to support GMCs.  On the other
hand, if $\eta\sim 1$ as suggested by the simulations
discussed in \S \ref{sec:mod}, then either winds are
much more efficient at energizing clouds ($\phi_w\gg 1$)
or most GMCs are energized by some other source, such
as massive stars.

   The concept of dynamically regulated star formation
can be applied to individual star--forming clumps as well.
Consider the star--forming clumps 
associated with the young star clusters
NGC 2023, 2024, 2068, 2071 in Orion B
\cite{lad91a} \cite{lad91b} \cite{lad99}.  The clumps
all have masses of order 400 \sun, although
the number of young stars associated with
the clumps ranges from 21 in NGC 2023 to 309 in
NGC 2024.    With a mean stellar mass
of 0.5 \sun\ \cite{sca86}, $\alpha\sim 1$,
$\Nbhtt\sim 1.5$, the dynamically regulated
star formation rate in one of these clouds
is $\dot\caln_{*\dr}\simeq 32/(\phi_w\eta)$ stars Myr\e.
Because of the small masses of the clumps involved,
large fluctuations in the star formation rate
may be expected.  Furthermore,
$\phi_w$ might be significantly less than unity
if the outflows can escape from the clumps \cite{mat99},
as has been observed in some cases \cite{rei97}.
Thus, the dynamically regulated
rate appears to be within the range observed in these
clumps assuming that the star formation has
been occurring for the past 1-3 Myr.
The OB star--forming clumps observed by Plume et al
\cite{plu97} are an order of magnitude more massive
than these, and it is not known if their properties are also
consistent with dynamically regulated star formation.

\subsection{PHOTOIONIZATION--REGULATED STAR FORMATION}

	To this point, we have discussed the star formation
rate in terms of how many stars must form in order to
provide adequate energy input to prevent molecular clouds
from collapsing, but we have not discussed the physical
mechanism that actually determines the rate at which
stars form.  For low mass stars, this time scale is believed
to be controlled by ambipolar diffusion \cite{mes56}
\cite{mou87} \cite{shu87}.  As we have seen in \S \ref{sec:mag2},
the magnetic critical mass for clumps in molecular clouds
is larger than the typical mass of a star.  In order
for gravity to overcome the force due to the magnetic 
field, gas must either accumulate along the field lines
or across the field; the latter can occur only via
ambipolar diffusion.  If the gas is to accumulate
along the field, then for $M/\mphi\simeq 2$ (\S \ref{sec:clo})
about half the mass along a given flux tube would have
to be concentrated into a single star.  A model 
in which the accumulation
of gas along flux tubes is regulated by wave damping
has been proposed recently by Myers \& Lazarian \cite{mye98}.
Here we shall focus on the role of ambipolar diffusion
in low mass star formation.

	  Most of the gas in a molecular cloud is neutral,
and does not interact directly with the magnetic field.
On the other hand, at typical densities in molecular clouds,
the charged particles are well coupled to the field.
As a result, if the neutral gas attempts to collapse under its
own self gravity, it will be restrained by collisions
with the charged particles.  Balancing the force of
gravity against that of friction gives
\beq
\frac{GM\rho}{R^2} \simeq n_i\langle \sigma v\rangle\rho\vad,
\eeq
where $\vad$, the ambipolar diffusion velocity, is the
relative velocity between the ions and the neutrals.
The time scale for ambipolar diffusion 
then depends only on the ionization of the gas \cite{spi78},
\beq
\tad\simeq \frac{R}{\vad} =\left(\frac{3\langle\sigma v\rangle}{4\pi
	   G\mu_{\rm H}} \right)x_e.
\eeq

	The numerical coefficient in the relation between $\tad$ and
$x_e$ depends on the geometry of the cloud and the nature
of the non--magnetic forces.  
To be specific, we identify $\tad$ as the
time for ambipolar diffusion to initiate
the formation of a very dense core.  
Fiedler \& Mouschovias \cite{fie93} simulated an 
axisymmetric cloud in which the thermal pressure
slowed the ambipolar diffusion significantly; their results
give $\tad/x_e\simeq 0.8\times 10^{14}$ yr.
Ciolek \& Mouschovias
\cite{cio94} found a similar result for the case
of a thin disk, $\tad/x_e\simeq 1.0\times 10^{14}$ yr.
We adopt
\beq
\tad=1.0\times 10^{14} \phad x_e~~~~{\rm yr},
\label{eq:tad}
\eeq
where $\phad$ is a constant of order unity that allows
for deviations from the typical value.  
In gas that is shielded from FUV radiation, the ionization
is due to cosmic rays (eq. \ref{eq:xe}), which implies
\beq
\frac{\tad}{\tff} = 23\phad \left(\frac{C_i}{10^{-5}~\mcm^{-3/2}}
		  \right).
\label{eq:tadtff}
\eeq
For $C_i\sim 0.7\times 10^{-5}$ (\S \ref{sec:che}),
$\tad\simeq 15 \tff$, so that
ambipolar diffusion is quite slow compared
to gravitational collapse.  

      When the magnetic critical mass $M_B$ significantly exceeds
the typical stellar mass, as appears to be the case
in Galactic molecular clouds (\S \ref{sec:mag2}), then
the rate at which stars form is set by the 
average rate of ambipolar diffusion in the cloud \cite{mck89}:
\beq
\dot M_*=\frac{M}{\tgs} \simeq\int\frac{dM}{\tad(x_e)} .
\eeq
The ionization in 
the outer parts of molecular clouds is relatively high
due to photoionization of the metals.  Regions in the clouds that are
shielded by an extinction $A_V>A_V$(CR)$\simeq 4$ are ionized 
primarily by cosmic rays; as a result, 
they have a lower ionization (eq. \ref{eq:xe})
and correspondingly lower ambipolar diffusion time.
As a result,
the star formation rate in a molecular cloud
is approximately that in the cosmic ray ionized region,
\beq
\dot M_*\simeq \frac{M(A_V>A_{\rm CR})}{\tad(x_{e{\rm CR}})} .
\eeq
Since the rate at which stars form is governed by
the mass that is shielded from photoionizing
FUV radiation, McKee termed this
{\it photoionization--regulated star formation}
\cite{mck89}.  This process is
naturally inefficient, both because only a fraction
of the cloud is in the cosmic ray ionized region
(models give a typical value of about 10\% \cite{mck89})
and because the ambipolar diffusion time
is much greater than the free fall time (eq. \ref{eq:tadtff}).

   How does this rate compare with that observed in
the Galaxy?  If we adopt $\nh\simeq 3000$
\cc\ from \S \ref{sec:dyn}, we find $\tad\simeq 2.4\times 10^7$ yr.  For
10\% of the mass in cosmic ray ionized regions, this gives
$\tgs\simeq 2.4\times 10^8$ yr.  Since the total
mass of molecular gas inside the solar circle
is about $1.0\times 10^9$ \sun\ \cite{wil97},
the predicted rate of photoionization regulated
star formation there is $\dot M_*=10^9\msun/2.4\times 10^8$ yr
$\simeq 4$ \sun\ yr\e, quite close to the observed value
of 3 \sun\ yr\e.

   One of the key predictions of this theory is
that star formation is restricted to regions
of relatively high extinction, $A_V\simgt A_{\rm CR}\simeq
4$ mag.  This has been tested in a study of the 
L1630 region of the Orion molecular cloud, and
indeed all the star formation was found to
be concentrated in regions in which the extinction
was greater than this.  McKee \cite{mck89}
also predicted that molecular clouds in the
Magellanic Clouds would have 
comparable extinctions, and therefore
higher column densities,
than Galactic molecular clouds in order
that star formation be able to provide the energy
needed to prevent the molecular clouds from
collapsing.  
Pak et al \cite{pak98} have found that the
column densities of molecular clouds in the LMC and SMC
do in fact scale approximately inversely with the metallicity,
consistent with an approximately constant extinction.

\section{Conclusion}

	We have seen that it is possible to understand
a number of the observed properties of molecular clouds
in terms of a model
in which GMCs and the clumps within them are modeled
as clouds in approximate hydrostatic equilibrium,
with the turbulence treated as a separate pressure
component.  In particular, we have seen why 
the clouds are generally gravitationally bound;
why they have approximately constant column densities;
why they have line widths that increase with size;
why, along with the star--forming clumps and cores
within them, they are somewhat 
magnetically supercritical; 
why they are the sites
of star formation in the Galaxy;
and why star formation
is inefficient.  Of course, this model is a drastically oversimplified
picture of the real situation.  Other researchers, looking
at the same data, have developed completely different
models:  For example, Elmegreen \& Falgarone \cite{elm96}
have developed a fractal model for structure
in the interstellar medium in which the concept
of pressure plays no role, and Ballesteros--Paredes et al
\cite{bal99} have argued that pressure balance is 
irrelevant in a turbulent medium such as the ISM.  By the time
of the next Crete meeting, there may be
some synthesis between these differing viewpoints
or perhaps an entirely new idea.  The challenge
facing us is formidible, for we must attempt to extend
our understanding to regions of OB star formation
as well.

\section{Acknowledgments}

I am deeply grateful to both Nick
Kylafis and Charlie Lada for providing me the opportunity
to learn more about star formation in such a stimulating
environment.  My thanks to Dick Crutcher for sharing his
results prior to publication, and to Frank Bertoldi,
Charles Curry, Chris Matzner, and Dean McLaughlin for
comments on the manuscript.  My work is supported in 
part by NSF grant AST95-30480, a grant from the
Guggenheim Foundation, a NASA grant that supports
the Center for Star Formation Studies, and
a grant from the Sloan Foundation to the Institute
for Advanced Study.

\def\aj{{\it AJ}}
\def\apj{{\it ApJ}}
\def\apjl{{\it ApJ Letters}}
\def\apjs{{\it ApJ Supp.}}
\def\aanda{{\it A\&A}}
\def\mnras{{\it MNRAS}}

\end{document}